\documentclass[%
 reprint,
 amsmath,amssymb,
 aps,
pre,
nolongbibliography
]{revtex4-2}

\usepackage{graphicx}
\usepackage{xcolor}
\usepackage{soul}
\usepackage{dcolumn}
\usepackage{bm}
\usepackage[colorlinks=true, allcolors=blue]{hyperref}

\usepackage{amsmath} 
\usepackage[utf8]{inputenc} 
\usepackage{physics}
\usepackage[letterpaper, margin=1in]{geometry}

\newcommand{\ord}{\mathcal O}

\newcommand{\figsubref}[2]{Fig.~\hyperref[#1]{\ref*{#1}(#2)}}

\begin{document}

\preprint{APS/123-QED}

\title{Accurate simulation of pulled and pushed fronts in the nonautonomous Fisher-Kolomogorov-Piskunov-Petrovsky equation}

\author{Troy Tsubota$^{1,2}$}
\email{troytsubota@g.harvard.edu}
\author{Smridhi Mahajan$^1$}
\author{Adrian van Kan$^{3,4,1}$}
\author{Edgar Knobloch$^1$}
\affiliation{$^1$Department of Physics, University of California at Berkeley, Berkeley, California 94720, USA}
\affiliation{$^2$Department of Physics, Harvard University, Cambridge, Massachusetts 02138, USA}
\affiliation{$^3$Institute of Geophysics, University of Münster, 48149 Münster, Germany}
\affiliation{$^4$Department of Mathematics, Texas A\&M University, College Station, Texas 77843, USA}

\date{October 8, 2026}

\begin{abstract}
We introduce a novel numerical method for direct simulation of front propagation in the Fisher-Kolomogorov-Piskunov-Petrovsky (F-KPP) equation with a time-dependent parameter on an infinite domain. The method computes a time-dependent boundary condition that accurately captures the leading-edge dynamics by coupling the nonlinear simulation region to a linear approximation region in which the dynamics can be solved exactly via the Green's function of the linearized equation. This approach enables precise front velocity measurements on relatively small computational domains for a variety of nonautonomous regimes and initial conditions for which existing numerical methods break down. We apply the method to pulled and pushed fronts in the F-KPP equation with quadratic and quadratic-cubic nonlinearities, finding that, in all cases, it significantly improves the accuracy of the simulated front velocity compared with a na\"ive treatment of boundary conditions, even for constant parameters and a fixed domain size. For pulled fronts with a diffusion coefficient that increases algebraically in time, our results reveal a deviation from the natural asymptotic velocity predicted by linear theory, whose explanation requires nonlinear theory. For pushed fronts with constant parameters, the method reproduces the exponential convergence to the theoretical asymptotic front speed and profile with improved precision. For a slowly time-varying linear growth parameter, we find that the pushed front velocity follows the changing parameter adiabatically if the asymptotic pushed velocity remains faster than the natural asymptotic pulled velocity. As the growth parameter moves toward the pushed--pulled transition point, the competition between the pushed and pulled fronts can result in both delayed and even premature onset of the pushed--pulled transition, depending on the form of parameter growth. The numerical method presented here proves to be an effective tool enabling useful new insights into front propagation in nonautonomous systems by reducing finite domain size effects and improving accuracy.
\end{abstract}


\maketitle

\section{Introduction}\label{sec:introduction}

In this article, we analyze the one-dimensional Fisher-Kolmogorov-Petrovsky-Piskunov (F-KPP) equation \cite{fisher_wave_1937,kolmogoroff_study_1988} with time-dependent parameters:
\begin{equation}
    u_t = d(t)u_{xx} + f(u,t), \label{eq:f-kpp}
\end{equation}
where $f(0,t)=0$, $f(1,t)=0$, $f_u(0,t)>0$, $f_u(1, t)<0$ and the subscripts denote partial derivatives. Thus, $u=0$ is a linearly unstable state, and the linearly stable state $u=1$ propagates into it. Following \cite{van_saarloos_front_2003}, we relax the classical KPP condition that $f_u(u,t) < f_u(0,t)$ for $u\in (0,1)$, which is imposed in the usual definition of a KPP-type equation \cite{kolmogoroff_study_1988,berestycki2022asymptotic}.

The F-KPP equation is the prototypical example of an equation that exhibits front propagation into an unstable state. Such front propagation, in which one pattern state invades another, is a ubiquitous phenomenon that arises in many physical and biological environments; see \cite{van_saarloos_front_2003} for an extensive review of applications. In general, there are two types of fronts. \textit{Pulled} fronts are governed by linear dynamics at their leading edge \cite{van_saarloos_front_2003}, although in some cases, their properties are not determined by the leading edge alone \cite{avery2025pulled}. \textit{Pushed} fronts are governed by nonlinear dynamics in the bulk of the front \cite{van_saarloos_front_2003}. Depending on the type and strength of the nonlinearities present, the F-KPP equation can exhibit both pulled and pushed fronts. Of great interest is the \textit{front selection} problem \cite{avery2025selection} for which we seek to determine the selected front velocity and profile given an initial condition, e.g., a localized perturbation. 

Front selection in the autonomous F-KPP equation has been studied rigorously in \cite{aronson_nonlinear_1975, aronson_multidimensional_1978, Bramson1983convergence}. Subsequent work established the \textit{marginal stability criterion} \cite{van_saarloos_front_2003, avery_universal_2022, avery2025selection} for pulled fronts, which states that for sufficiently steep initial conditions, the front approaches the marginally stable front solution with asymptotic velocity and steepness determined by linear dynamics. This criterion has been shown to hold for a wide class of equations \cite{van_saarloos_front_2003, avery_universal_2022,klikaSpeedPropagationTuring2024} among other related ideas such as marginal positivity \cite{van_saarloos_front_2003, avery2025selection}. Pushed fronts in the autonomous case have also been well-studied, giving rise to the concept of \textit{nonlinear marginal stability}, although analyses are specific to the choice of nonlinearity \cite{rothe1981convergence,van1989front,van_saarloos_front_2003}.

On the other hand, front selection in the nonautonomous case is not as well studied, and its understanding would provide insight into important problems such as pattern formation on growing domains \cite{knobloch_problems_2015,GohScheel2023,tsubota_bifurcation_2024}, relevant to the growth of organisms, and ecological systems within a time-varying environment \cite{ducrot_spreading_2023,alzaleq_analysis_2020}. Previous studies \cite{tsubota_bifurcation_2024,AVERY2025134972,mendez_speed_2003,nadin_propagation_2012,berestycki2022asymptotic} have made some progress on the nonautonomous front selection problem, particularly for the F-KPP equation. Pulled fronts in the nonautonomous F-KPP equation admit a \textit{natural asymptotic velocity} and \textit{natural asymptotic steepness}, which differ from the velocity and steepness expected from the marginal stability criterion \cite{tsubota_bifurcation_2024}. However, these studies are restricted to limited parameter regimes. Furthermore, while the natural asymptotic velocity and steepness have been shown to be useful quantities in some cases, there are currently no general guarantees that the front asymptotically converges to this velocity and steepness. Nonautonomous pushed fronts also remain largely unexplored.

To make progress towards a more general theory, we turn to direct numerical simulations. These pose two key challenges: (i) we are interested in asymptotic dynamics on an infinite domain but can only simulate finite ones, and (ii) small amplitude instabilities ahead of a pulled front may impact the measured front speed, cf.~\cite{ponedel_front_2017}. Simulations on domains of modest length are desirable because they reduce the available time for the growth of these instabilities and the required computational resources. However, such simulations require great care in the formulation of the downstream boundary conditions in order to avoid altering the properties of the solution. The challenge posed by the finiteness of the domain is typically addressed by making the domain large and enforcing what one may consider reasonable boundary conditions, e.g., zero Dirichlet or zero Neumann. However, for pulled fronts, the behavior at the leading edge of the front ($x\to\infty$) is critical \cite{van_saarloos_front_2003, ebert_front_2000}, and na\"ive boundary conditions do not capture the dynamics accurately. For example, a small-amplitude cut-off is known to shift the front speed, among other effects \cite{brunet1997shift,panja2002weakly}.

These challenges are somewhat mitigated by going to a frame that moves with the front \cite{avery2025selection}. Heuristically, a comoving frame allows the front to remain away from the simulation boundaries for longer times, allowing for longer simulation times with smaller computational domains. However, the challenges described above are not entirely eliminated. Moreover, nonautonomous problems like the F-KPP problem (\ref{eq:f-kpp}) in which the front may continue to evolve as it propagates, pose a nontrivial question as to how a comoving frame should be defined. Additionally, even for the autonomous problem, the importance of the downstream boundary condition is only amplified. A moving frame alone does not significantly improve the simulation accuracy at long times because the effect of the na\"ive boundary condition propagates into the interior of the computational domain \cite{ebert_front_2000}.



\begin{figure}[!htbp]
\includegraphics[width=0.48\textwidth]{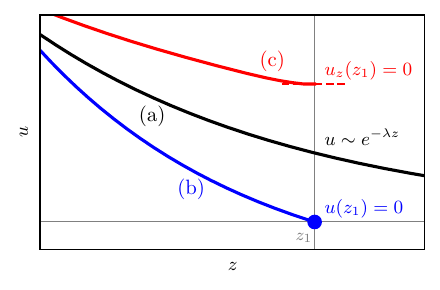}
\caption{Na\"ive choices for the boundary condition at the edge of the moving frame $z=z_1$ do not accurately capture the dynamics at the leading edge of a front in the F-KPP equation on the real line. (a) The exponential leading edge of a front with steepness $\lambda$. A desirable ``infinity-simulating'' boundary condition would give correct values for $u(z_1)$ and $u_z(z_1)$ of this tail, where $z=z_1$ is the location of the boundary. (b) A zero Dirichlet boundary condition $u(z_1)=0$ results in a front that is too steep and therefore too slow. (c) A zero Neumann boundary condition $u_z(z_1) = 0$ results in a front that is too shallow and therefore too fast.}
\label{fig:boundary_condition}
\end{figure}

To examine this issue more closely, we go into a moving frame with coordinate $z$ and refer to Fig.~\ref{fig:boundary_condition}. The leading edge of a front in the F-KPP equation generally has an exponential tail $u(z,t) \sim e^{-\lambda z}$ as $z\to\infty$, where $\lambda$ is the steepness [\figsubref{fig:boundary_condition}{a}]. Roughly speaking, for a pulled front, the steepness dictates its speed: if the front is steeper, it moves more slowly. A good boundary condition would faithfully represent this exponential tail. However, neither a zero Dirichlet condition $u(z_1)=0$ nor a zero Neumann condition $u_z(z_1)=0$ can satisfy this requirement. For a zero Dirichlet condition [\figsubref{fig:boundary_condition}{b}], pinning the front at zero prevents it from developing into its desired exponential tail structure. The resulting front propagates at a systematically reduced speed \cite{ebert_front_2000,ponedel_front_2017}, which can be particularly extreme for shallow fronts. Conversely, a zero Neumann boundary condition [\figsubref{fig:boundary_condition}{c}] forces the leading edge to be shallower than it should be, resulting in a faster propagating front. In fact, the Neumann condition creates a flat state in long times that expands uniformly \cite{ebert_front_2000}. A Robin boundary condition may seem to provide a proper mix of the Dirichlet and Neumann conditions. In fact, for steepness $\lambda$, the Robin condition $u_z + \lambda u = 0$ would represent this exponential tail precisely. However, this technique only works for shallow initial conditions. Steep initial conditions approach the asymptotic steepness from above, so enforcing this Robin condition starting at $t=0$ would impose an incorrect constraint at early times. This technique also requires a priori knowledge of the front steepness that we do not have in the nonautonomous case. A related approach accounts for the exponential tail by setting a nonzero Dirichlet boundary condition with an exponential form \cite{zhao_comparison_2003}, yet similar problems arise.

The problem of simulating an infinite domain on a finite computational domain has been studied for other equations \cite{givoli1999recent,givoli2013numerical}; it is related to the problem of non-reflecting boundary conditions, also known as perfectly matched layers, in wave propagation theory \cite{keller1989exact,Givoli2008}. In particular, for equations that reduce to the heat equation in the far-field, \cite{ditkowski_near-field_2008} introduces a Green’s function approach to construct exact boundary conditions. The method analytically eliminates the exterior infinite domain and replaces it with a Neumann–to–Dirichlet relation that prescribes the correct boundary value by integrating over the flux into the exterior region. The study also introduces the notion of a buffer region between the interior and exterior region so that the resulting convolutions with the Green's function are well-behaved and can be reliably computed numerically. An earlier version of this approach was applied to the Fisher equation in \cite{hagstrom_numerical_1986} using a Laplace transform on the linearized equation outside of the domain. However, arbitrary time-dependent parameters are difficult to incorporate into the Laplace transform formulation.

Here, we combine the insights from previous work and apply them to the nonautonomous problem. We shift into a moving frame and introduce an infinity-simulating boundary condition built on the exact Green's function associated with the linearized F-KPP equation that accurately represents the leading-edge dynamics. We refer to this method as the Green's (function) boundary condition (GBC) method in the following. This approach removes spurious boundary effects and allows reliable long-time front dynamics to be computed accurately on much smaller domains, as demonstrated below for a variety of time-dependent regimes in both pulled and pushed settings. We focus on two key cases: the nonautonomous F-KPP with a quadratic nonlinearity, i.e. the Fisher equation, and the nonautonomous F-KPP with a quadratic-cubic nonlinearity.

Throughout the paper we compare our results with the ``time-frozen" case, i.e. the case in which the time variable in the coefficients is treated as a parameter and the problem becomes autonomous and hence easily solvable. This terminology differs from that of Beyn et al.~\cite{beyn2004,beyn2008} where group orbits of solutions are computed by evolving a ``frozen" state in time.

\section{Numerical Approach}\label{sec:numerical}

\begin{figure*}
\centering
\includegraphics[width=\textwidth]{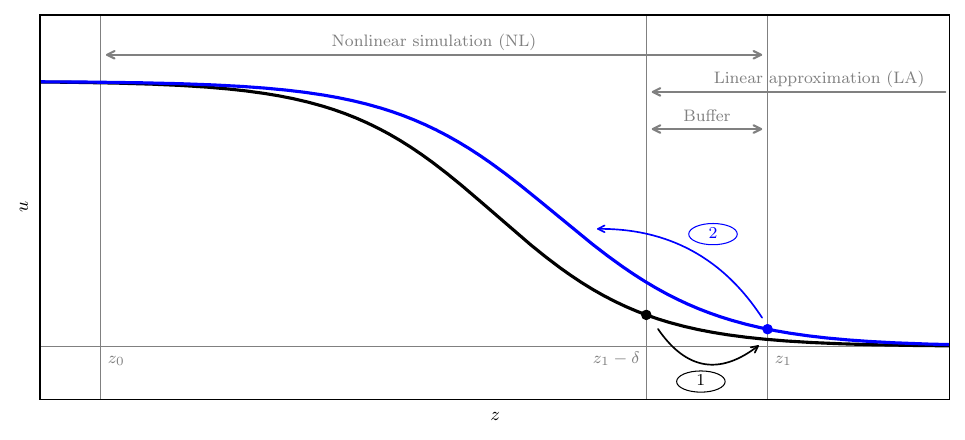}
\caption{In the Green's (function) boundary condition (GBC) method, the domain is split into two regions: the nonlinear (NL) simulation region and the linear approximation (LA) region. These regions, defined with respect to the moving frame coordinate $z$, overlap in a buffer region of size $\delta$ required for regularization. At each time step, (1) the value $u(z_1-\delta, t)$ acts as a boundary forcing for the LA region to predict $u(z_1, t+dt)$. Then, (2) the updated value $u(z_1, t+dt)$ is the new boundary condition for an implicit step to compute $u(z,t+dt)$ within the NL region. The left boundary condition is a zero-flux condition $u_z(z_0, t)=0$.}
\label{fig:numerical_approach}
\centering
\end{figure*}

In the general case, we seek to simulate Eq.~\eqref{eq:f-kpp} on an infinite domain $x\in(-\infty, \infty)$. We shift into a moving frame with a prescribed velocity $c(t)$ (not necessarily comoving with the front) to obtain
\begin{equation}
    u_t = d(t)u_{zz} + c(t)u_z + f(u,t),
\end{equation}
where $z=x-\int_0^t c(t')dt'$ is the moving frame coordinate. The key idea behind the GBC 
method is that \textit{the leading edge is governed by linear dynamics.} We can select $c(t)$ to be such that the leading edge values remain small for the entire simulation time. For instance, the comoving velocity would satisfy this requirement.

Now, refer to Fig.~\ref{fig:numerical_approach} for the high-level approach. The domain can be decomposed into two regions: the nonlinear simulation (NL) region on $[z_0,z_1]$ and the linear approximation (LA) region on $[z_1-\delta, \infty)$ with $\delta>0$. The NL and LA regions communicate through their boundaries. More precisely, suppose we begin with a compactly supported initial condition $u(z,0)=u(x,0)$ on $[z_0,z_1-\delta]$, so that $u$ is initially identically zero in the LA region. Then, the NL region quantity $g(t) \equiv u(z_1-\delta, t)$ provides  boundary forcing for the LA region and completely determines the solution in the LA region. This can be formulated as the problem
\begin{align}
    u_t &= a(t)u + c(t)u_z + d(t)u_{zz}, \\
u(z,0) &= 0, \ \ z > z_1-\delta, \\
    u(z_1-\delta,t) &= g(t), \ \ t>0,
\end{align}
to be solved in the LA region $z\in[z_1-\delta, \infty)$. The function $a(t)$ arises from the linearization of $f(u,t)$ around $u=0$. Given a boundary Green's function $G_b(z,t;y,s)$ for this equation (for the heat equation, see \cite{carslaw1959}), we can write the solution in the LA region as
\begin{equation}
    u(z,t) = \int_0^t G_b(z,t;z_1-\delta,s)g(s)ds.\label{eq:green_function_convolution_simple}
\end{equation}
In particular, we can compute $u(z_1,t)$. Conversely, we can use the value $u(z_1,t)$ as a time-dependent Dirichlet boundary condition for the NL region.

The detailed computation of the Green's function is rather involved; see Appendix~\ref{app:theoretical} for the complete treatment. Importantly, we are able to obtain closed-form expressions for the Green's function and its convolution integrals under the key assumption that $c(t)=\gamma d(t)$ for some constant $\gamma$. This assumption is not too restrictive since we can tune $\gamma$ as desired, but it is important to recognize that we cannot necessarily move at the comoving velocity. Additionally, the Green's function formulation allows for a straightforward generalization to nonzero initial conditions in the LA region.

The GBC method proceeds as follows (Fig.~\ref{fig:numerical_approach}). We discretize time with a time step $dt$. At each time step, we perform the following two steps:
\begin{enumerate}
    \item Use the NL region value $g(t) \equiv u(z_1-\delta, t)$ to compute $u(z_1, t+dt)$ in the LA region by convolving with the boundary Green's function.
    \item Use $u(z_1, t+dt)$ as a Dirichlet boundary condition to compute $u(z,t+dt)$ from $u(z,t)$ in the NL region by a numerical time step.
\end{enumerate}

For our simulations, which are systematically performed in the moving frame, we use second-order centered finite differences in space and an implicit-explicit (IMEX) Euler scheme in time, which treats the linear terms $a(t)u + c(t)u_z + d(t)u_{zz}$ implicitly and the nonlinear terms in $f(u,t)$ explicitly. We implement a simple Neumann boundary condition $u_z(z_0,t) = 0$ at the left boundary; since the system saturates at the stable state $u=1$ there, this boundary does not significantly affect the front dynamics in the F-KPP equation. However, one could implement a similar linear approximation for the left boundary by linearizing around the stable $u=1$ state in the region $(-\infty, z_0]$.

Compared to a typical finite-difference solver, several new numerical considerations must be taken into account, such as the evaluation of the convolution integrals, the size of the buffer region, and the speed of the frame. These details, which are not important for understanding the main results, are discussed in Appendix~\ref{app:numerical}.

\section{Pulled fronts}\label{sec:pulled}

\subsection{Autonomous pulled fronts}

\begin{figure*}
\centering
\includegraphics[width=\textwidth]{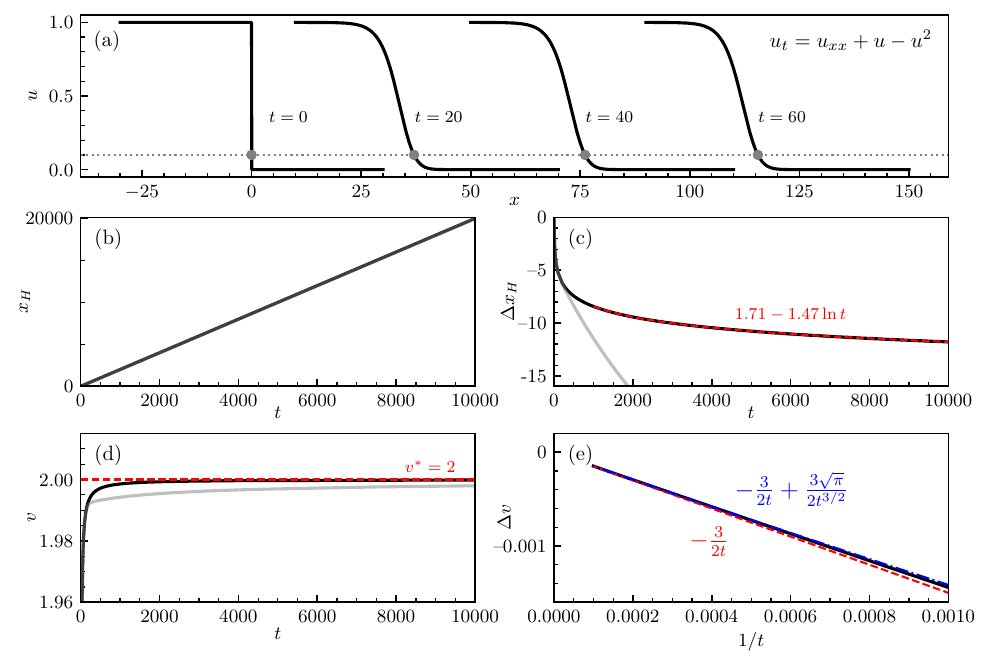}
\caption{The GBC method benchmarks well for the simplest case of the autonomous Fisher equation \eqref{eq:fisher_time_independent} with a step function initial condition. The GBC simulations use $dt=0.025$, $t\in[0,10\,000]$, $dz=0.01$, $z\in[-30, 30]$, where the variable $z\equiv x-ct$ refers to the frame moving with the asymptotic speed $c=v^*=2$. A buffer region of size $\delta = 10$ is used for the GBC nonlinear-linear transition (see Fig.~\ref{fig:numerical_approach}). (a) Lab-frame visualization of the propagating front. The solid, black curves at $t=0$, $t=20$, $t=40$, and $t=60$ are the numerically simulated profiles obtained with the GBC method. The dotted, gray line with solid points indicates the tracked front position at $H=0.1$. (b) Front position $x_H(t)$ in the lab frame with the GBC method (solid, black) and a na\"ive Dirichlet zero (DZ) condition (solid, gray). The difference in this plot is not perceptible. Note that DZ simulations use a domain $z\in[-90,30]$ to keep the front within the simulated domain. (c) Front position $\Delta x_H(t) = x_H(t) - v^*t$ in the moving frame for the two cases.  Here, we clearly see that the GBC front correctly follows the theoretical $\ln t$ scaling (fitted with a red dashed line), while the DZ front does not. (d) Front velocity $v(t)=v_H(t)$ in the lab frame, computed by averaging over a time window $\Delta t = 5 = 200\,dt$. The asymptotic value $v^*=2$ (dashed, red) is approached from below in the GBC method, but the velocity undershoots in the DZ method. This slowdown in the DZ case is expected (cf. Fig.~\ref{fig:boundary_condition}). (e) Front velocity deviation from its asymptotic value, $\Delta v_\mathrm{nlin}(t) \equiv v(t) - v^*$, plotted over $1/t$, approaches the $-3/2t$ line (dashed, red) as $t\to\infty$, in agreement with theory \cite{Bramson1983convergence,van_saarloos_front_2003,avery_universal_2022}. Further, the measured velocity agrees well with the $\ord(t^{-3/2})$ correction (dash-dotted, blue), improving upon previous measurements (cf. \cite{ebert_front_2000}, Fig.~7).} 
\label{fig:constant_steep}
\centering
\end{figure*}

To benchmark the GBC method, we first consider the simpler autonomous Fisher equation
\begin{equation}
    u_t = u_{xx} + u - u^2.\label{eq:fisher_time_independent}
\end{equation}
It is well-known that for steep initial conditions, e.g. a step function, delta function, or Gaussian, the asymptotic velocity of a front in Eq.~\eqref{eq:fisher_time_independent} is $v^*=2$ via the marginal stability criterion or other methods. Furthermore, the asymptotic approach to this velocity is \cite{van_saarloos_front_2003, Bramson1983convergence, avery_universal_2022}
\begin{equation}
    \Delta v_\mathrm{nlin}(t) \equiv v(t) - v^* = -\frac{3}{2t} + \frac{3\sqrt{\pi}}{2t^{3/2}} + \text{h.o.t.}\label{eq:v_nlin_scaling}
\end{equation}
The leading $\ord(1/t)$ scaling can also be recovered from a simple linear argument as follows. The exact solution to Eq.~\eqref{eq:fisher_time_independent} linearized about $u=0$ with a delta function initial condition is
\begin{equation}
    u(x,t) = \frac{1}{\sqrt{4\pi t}}\exp\left(t - \frac{x^2}{4t}\right).
\end{equation}
Tracking the front at a height $u=H$ gives the exact position (in the lab frame)
\begin{equation}
    x_{H,\mathrm{lin}}(t) = 2t\left[1-\frac{1}{2t}\ln\left(4\pi H^2 t\right)   \right]^{1/2}. 
\end{equation}

We use this result to define the linear front speed $v_{H,\mathrm{lin}}(t)\equiv d x_{H,\mathrm{lin}}/dt$, and conclude that for large $t$ the velocity deviation
\begin{equation}
    \Delta v_{\mathrm{lin}}(t) \equiv v_{H,\mathrm{lin}}(t) - v^* =  - \frac{1}{2t} + \text{h.o.t.}
\end{equation}
While the $1/t$ coefficient differs between the linear and nonlinear cases \cite{van_saarloos_front_2003}, the $\ord(1/t)$ scaling is the same. In the following, we shall define, with a slight abuse of notation, other quantities called $\Delta v$, in each case referring to the deviation of the front velocity from the relevant asymptotic velocity.

We use the GBC method to numerically solve Eq. \eqref{eq:fisher_time_independent} with a step function initial condition $u(x,0) = 1$ for $x\leq 0$ and $u(x,0)=0$ for $x>0$. For comparison, we define the Dirichlet zero (DZ) method as the same method with the same parameters but without the dynamic boundary condition. We show the results in Fig.~\ref{fig:constant_steep}, which indicates that the GBC method performs well for a steep initial condition: the front correctly approaches $v^*=2$ from below with the predicted nonlinear scaling \eqref{eq:v_nlin_scaling}, while the DZ method results in an incorrect velocity that undershoots $v^*$. The DZ method forces the front to be too steep, as depicted in Fig.~\ref{fig:boundary_condition}, resulting in a slower front.

\begin{figure*}
\centering
\includegraphics[width=\textwidth]{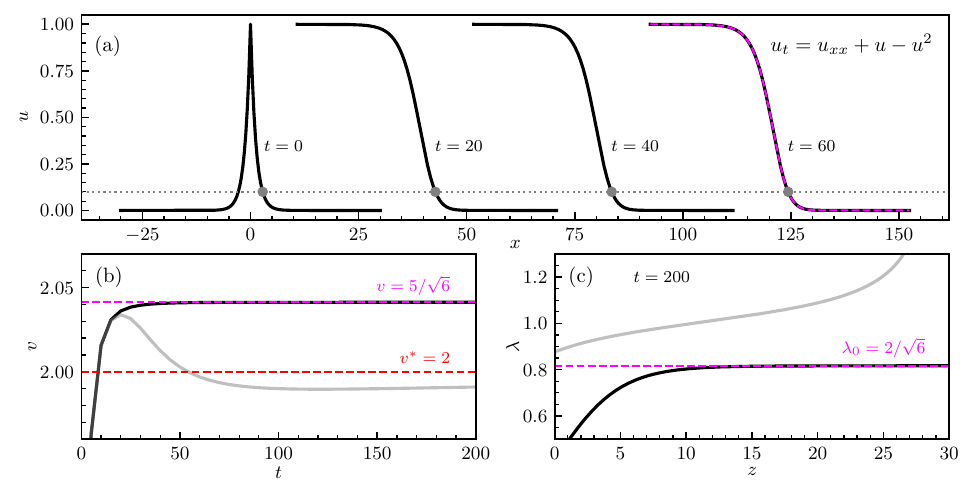}
\caption{The GBC method also benchmarks well for Eq.~\eqref{eq:fisher_time_independent} with an exponential initial condition $e^{-\lambda_0|x|}$. Here, $\lambda_0=2/\sqrt{6}$ which is less than the critical steepness $\lambda^*=1$. We set the moving frame speed $c$ to be the asymptotic speed $v=\lambda_0 + 1/\lambda_0 = 5/\sqrt{6}$. (a) Numerically simulated front (solid, black) with this initial condition propagates through the lab frame. This particular case has the known exact asymptotic solution given in Eq.~\eqref{eq:shallow_exact}. At $t=60$, the numerical solution shows good agreement with this exact solution (magenta, dashed). (b) Front velocity over time shows that the GBC front (solid, black) correctly attains the theoretical asymptotic velocity $v=5/\sqrt{6}$ (magenta, dashed), while the DZ front (solid, gray) eventually falls below $v^*=2$ (red, dashed). (c) Local steepness $\lambda = -u_z/u$ in the moving frame coordinate $z\equiv x-\int_0^t c(t')\,dt'$ at $t=200$ shows that the GBC front steepness correctly approaches $\lambda_0 = 2/\sqrt{6}$ at the leading edge while the DZ front does not.}
\label{fig:constant_shallow}
\centering
\end{figure*}

As described in \cite{van_saarloos_front_2003}, the meaning of a ``steep'' initial condition is that the tail as $x\to\infty$ falls off faster than $e^{-\lambda^* x}$, where $\lambda^*$ is the asymptotic steepness. For Eq.~\eqref{eq:fisher_time_independent}, one finds $\lambda^*=1$. An exponential initial condition with steepness $\lambda_0 < \lambda^*$ is considered a ``shallow'' (or ``flat'') initial condition \cite{ebert_front_2000}. In this case, the front does not approach the marginally stable front solution but instead converges to a uniformly translating solution with the same steepness $\lambda_0$ and speed $v = \lambda_0 + 1/\lambda_0 > v^*$ \cite{van_saarloos_front_2003}. For $\lambda_0=2/\sqrt{6}$, this front has a known exact solution (up to translation invariance) given by
\begin{equation}
    u^*(x,t) = \left\{1+\exp\left[\sqrt{\frac{1}{6}}\left(x-\frac{5}{\sqrt{6}}t\right)\right]\right\}^{-2}\label{eq:shallow_exact}
\end{equation}
with speed $v=5/\sqrt{6}$ \cite{zhao_comparison_2003}.

The GBC method unlocks the ability to accurately study this front profile. Under the DZ method, the front steepness diverges and profile measurements become increasingly unreliable as the edge of the domain is approached. The GBC method resolves this issue by assigning the correct value of the front at the boundary, which allows for the front to be well behaved here, enabling study of the profile even in small domains.


In Fig.~\ref{fig:constant_shallow}, we show that the GBC method also performs well for the shallow exponential initial condition $u(x,0) = e^{-\lambda_0|x|}$ with $\lambda_0=2/\sqrt{6}$. The GBC
front correctly approaches $v=5/\sqrt{6}$, and the profile agrees with the exact solution at long times. The DZ front eventually falls below $v^*=2$ because the na\"ive Dirichlet boundary condition does not faithfully represent the shallow leading edge, and the front regresses to the steep initial condition regime. 

Already in the autonomous case, the power of the GBC method becomes apparent. The method can handle both the steep and shallow cases within the same framework. This contrasts with previous methods that require different boundary conditions for each case.

\subsection{Nonautonomous pulled fronts}

\begin{figure}
\includegraphics[width=0.48\textwidth]{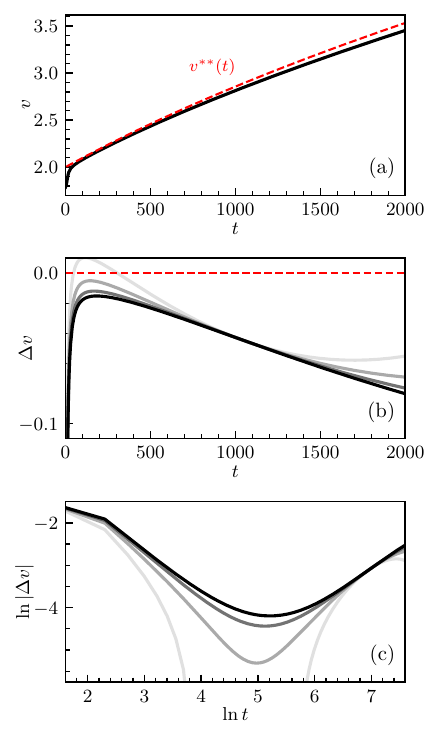}
\caption{The GBC method applied to the Fisher equation with linearly increasing time-dependent diffusion coefficient $d(t) = 1+\epsilon t$, $\epsilon = 0.001$. The numerical simulation parameters are $dt=0.025$, $t\in[0,2\,000]$, $dz=0.01$, $z\in[-100, 400]$ in a moving frame with speed $c(t)=\gamma d(t)$, $\gamma=1.40$, and a buffer region of size $\delta = 10$. (a) The numerical front velocity $v(t)$ (solid black) deviates from the natural asymptotic velocity $v^{**}(t)$ (dashed red) as $t\to\infty$ in disagreement with linear theory. (b) The velocity difference $\Delta v(t) \equiv v(t) - v^{**}(t)$ reveals this disagreement more clearly. Simulations with $dt\in \{0.2, 0.1, 0.05, 0.025\}$ are shown in progressively darker shades of gray, with the $dt=0.025$ simulation depicted in (a) drawn in black. (c) A log-log plot of $\Delta v(t)$ suggests the existence of short- and long-time asymptotic algebraic regimes.}
\label{fig:d_linear}
\end{figure}

\begin{figure}
\includegraphics[width=0.48\textwidth]{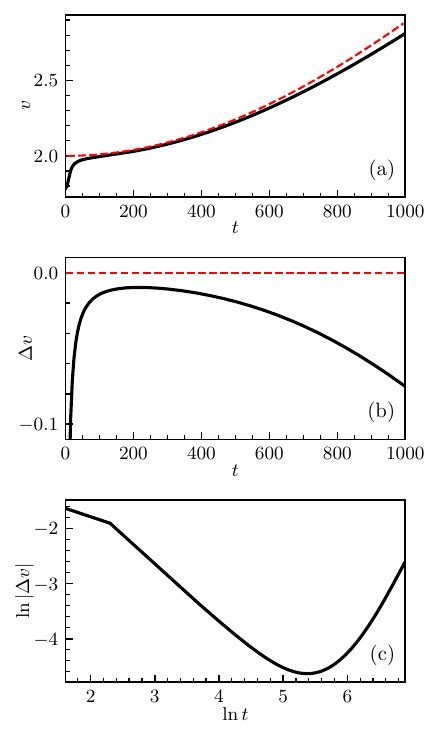}
\caption{As in Fig.~\ref{fig:d_linear}, a quadratically increasing diffusion coefficient $d(t) = 1+(\epsilon t)^2$ with $\epsilon = 0.001$ also indicates two-stage algebraic asymptotics.}
\label{fig:d_quadratic}
\end{figure}

\begin{figure}
\includegraphics[width=0.48\textwidth]{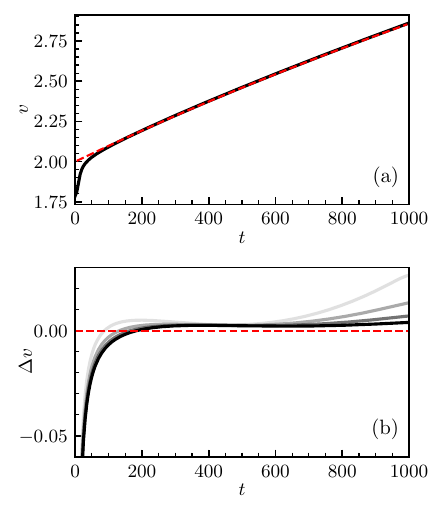}
\caption{In contrast to the linearly increasing diffusion coefficient (Fig.~\ref{fig:d_linear}), a linearly increasing growth coefficient $a(t) = 1+\epsilon t$ with $\epsilon = 0.001$ shows convergence as $dt$ is reduced, in agreement with the rigorous proof \cite{AVERY2025134972}. Line colors and the definition of $\Delta v$ are identical to those in Fig.~\ref{fig:d_linear}.}
\label{fig:a_linear}
\end{figure}

\begin{figure}
\includegraphics[width=0.48\textwidth]{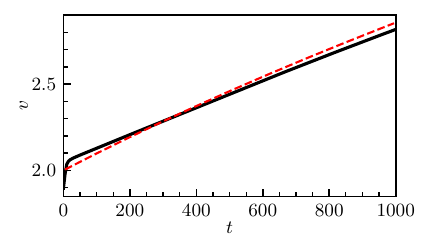}
\caption{The same linearly increasing diffusion coefficient $d(t) = 1+\epsilon t$ with $\epsilon = 0.001$ as in Fig.~\ref{fig:d_linear} but with an exponential initial condition $u(x,0) = e^{-\lambda |x|}$, $\ \lambda=0.9$. Initially, the front steepness is less than the natural asymptotic steepness, i.e. $\lambda < \lambda^{**}(0) = 1$, so the front propagates faster than the natural asymptotic velocity. However, since $\lambda^{**}(t)$ decreases and $v^{**}(t)$ increases with increasing $t$, the front velocity eventually falls below the natural asymptotic velocity. Line colors are identical to those in Fig.~\ref{fig:d_linear}.}
\label{fig:d_linear_exp_ic}
\end{figure}

Now we consider the nonautonomous Fisher equation
\begin{equation}
    u_t = a(t)u + d(t)u_{xx} - a(t)u^2\label{eq:fisher-basic}
\end{equation}
and linearize around $u=0$:
\begin{equation}
    u_t = a(t)u + d(t)u_{xx}.\label{eq:fisher-linearized}
\end{equation}
Using linear theory, one can deduce that the long-time behavior of pulled fronts in Eq.~\eqref{eq:fisher-basic} may be described by the \textit{natural asymptotic velocity} and \textit{natural asymptotic steepness} \cite{tsubota_bifurcation_2024}, defined as
\begin{align}
    v^{**}(t) &= \frac{A(t)d(t) + D(t)a(t)}{\sqrt{A(t)D(t)}}, \label{eq:natural_asymptotic_velocity} \\
    \lambda^{**}(t) &= \sqrt{\frac{A(t)}{D(t)}},\label{eq:natural_asymptotic_steepness}
\end{align}
where $A(t) \equiv \int_0^t a(t')\,dt'$ and $D(t) \equiv  \int_0^t d(t')\,dt'$. These differ from the time-frozen asymptotic velocity and steepness
\begin{align}
    v^{*}(t) &= 2\sqrt{a(t)d(t)}, \label{eq:time_frozen_asymptotic_velocity} \\
    \lambda^{*}(t) &= \sqrt{\frac{a(t)}{d(t)}} \label{eq:time_frozen_asymptotic_steepness}
\end{align}
obtained from the marginal stability criterion. Evidently, the marginal stability criterion does not adequately describe the asymptotic behavior of pulled fronts in the nonautonomous case. Hysteretic effects play an important role and are accounted for in Eqs.~\eqref{eq:natural_asymptotic_velocity} and \eqref{eq:natural_asymptotic_steepness}. For more details, see \cite{tsubota_bifurcation_2024}.

However, there are no general guarantees that the front velocity converges to $v^{**}(t)$ in the fully nonlinear setting, unless $a(t)$ grows linearly  \cite{AVERY2025134972} and in some cases of bounded parameters \cite{berestycki2022asymptotic}, and the authors are unaware of any general results on the convergence asymptotics in the nonlinear regime. In the autonomous case described above, the nonlinear leading-order asymptotics of the front velocity agree with the linear asymptotics up to a multiplicative constant. 
We follow the same procedure here to obtain a preliminary prediction for the rate of convergence of the nonautonomous front to its natural asymptotic velocity.

The exact solution to Eq.~\eqref{eq:fisher-linearized} for a delta function initial condition is
\begin{equation}
    \label{eq:fundamental_solution_nonautonomous_Fisher}
    u(x,t) = \frac{1}{\sqrt{4\pi D(t)}}\exp\left(A(t) - \frac{x^2}{4D(t)}\right).
\end{equation}
Solving for the front position at a tracking height $H$ gives the exact front position (in the lab frame)
\begin{equation}\label{eq:linear_front_position_exact}
    x_H(t) = 2\sqrt{A(t)D(t)} \left[ 1-\frac{1}{A(t)} \ln\left(H\sqrt{4\pi D(t)}\right) \right]^{1/2}.
\end{equation}
In order to proceed further, we restrict to asymptotically algebraic growth of $a(t)$ and $d(t)$ so that we can Taylor expand the second term in Eq.~(\ref{eq:linear_front_position_exact}) in the limit $t\to\infty$. This results in 
\begin{equation}\label{eq:linear_front_position_expansion}
    x_H(t) = x^{**}(t) - \sqrt{\frac{D(t)}{A(t)}} \ln\left(H\sqrt{4\pi D(t)}\right) + \text{h.o.t.},
\end{equation}
where $x^{**}(t) = \int_0^tv^{**}(s)\,ds = 2\sqrt{A(t)D(t)}$ is the \textit{natural asymptotic position} as described in \cite{tsubota_bifurcation_2024}.

Let $a(t)=\ord(t^{\alpha})$ and $d(t)= \ord(t^{\beta})$ as $t\to\infty$ for some $\alpha, \beta \geq 0$. Thus $A(t) = \ord(t^{\alpha+1})$ and $D(t) = \ord(t^{\beta+1})$. It follows that the deviation of the instantaneous front position from its natural asymptotic position scales as
\begin{equation}
    \Delta x(t) \equiv x_H(t) - x^{**}(t) = \ord\lbrace t^{(\beta-\alpha)/2}\ln t\rbrace.
\end{equation}
When $\beta-\alpha \neq 0$ the corresponding front velocity deviation scales as
\begin{equation}\label{eq:linear_asymptotics}
    \Delta v(t) \equiv v_H(t) - v^{**}(t) = \ord\left\{t^{(\beta-\alpha)/2 - 1}\ln t\right\},
\end{equation}
while for $\beta-\alpha = 0$ we have instead $\Delta v(t) = \ord(1/t)$ as in the autonomous case.

For example, for $a(t) = 1$ and $d(t) = 1 + \epsilon t$, we expect $\Delta v(t) \sim t^{-1/2}\ln t$ as $t\to\infty$. This suggests that the front velocity should converge to the natural asymptotic velocity $v^{**}(t)$ as $t\to\infty$. In general, based on Eq.~(\ref{eq:linear_asymptotics}), we expect convergence to the natural asymptotic velocity if $\beta-\alpha < 2$.

We can now use the GBC method to simulate Eq.~\eqref{eq:fisher-basic} and test these linear asymptotic predictions. For a linearly increasing diffusion coefficient $d(t) = 1 + \epsilon t$ with $\epsilon = 0.001$ and a constant growth coefficient $a(t)=1$, the results are shown in Fig.~\ref{fig:d_linear}. We see that, for a step function initial condition, the front velocity does not converge to the natural asymptotic velocity, contrary to the predictions of linear theory. In fact, when examining $\ln |\Delta v|$ over $\ln t$ as in \figsubref{fig:d_linear}{c}, the asymptotic behavior is quite complex. Initially, the velocity deviation appears to decrease algebraically in a transient regime. However, at long times, the velocity deviation appears to increase algebraically. We understand this behavior as follows. Over an initial period, $0\leq t\ll\epsilon^{-1}$, the coefficient $d(t)$ hardly changes (provided $d(0)>0$) and the system therefore resembles the autonomous case with $d=d(0)$ and so the front speed evolves towards the natural asymptotic speed. Once $t\gg\epsilon^{-1}$ the system senses the full time dependence and the front velocity begins to diverge from the predicted speed. This two-stage algebraic behavior is more pronounced for the quadratically increasing diffusion coefficient $d(t) = 1 + (\epsilon t)^2$ as shown in Fig.~\ref{fig:d_quadratic}, although in this case linear theory does not predict convergence.

To confirm that this is not an artifact of the numerical method used, we perform a range of simulations with $dt\in\{0.2,\,0.1,\,0.05,\,0.025\}$ as shown in \figsubref{fig:d_linear}{b}. The deviation from the natural asymptotic velocity persists as $dt$ is reduced, indicating that this is indeed a physical effect rather than a discretization error. Additionally, in Fig.~\ref{fig:a_linear}, we show that a linearly increasing growth coefficient $a(t) = 1 + \epsilon t$ with constant diffusion ($\alpha=1$, $\beta=0$) converges to the natural asymptotic velocity as proven in \cite{AVERY2025134972}. Comparing \figsubref{fig:constant_steep}{d} with \figsubref{fig:a_linear}{b} and relying on the intuition provided by linear theory \eqref{eq:linear_asymptotics}, our results suggest that a monotonically increasing $a(t)$ enables faster convergence to the natural asymptotic velocity than in the autonomous case. 


Note that the shapes of the curves in Fig.~\ref{fig:d_linear} are also influenced to some extent by the manually selected speed of the frame. However, an appropriate choice of frame speed does not influence the overall result. See Appendix~\ref{app:numerical} for further commentary on the choice of the frame speed parameter $\gamma$.

Another effect we observe with the GBC method is that of a time-dependent parameter on a shallow initial condition. For example, for $d(t) = 1 + \epsilon t$ and an exponential initial condition $u(x,0) = e^{-\lambda|x|}$ with $\lambda=0.9< \lambda^{**}(0) = 1$, Fig.~\ref{fig:d_linear_exp_ic} shows that, as expected, the front velocity initially exceeds the natural asymptotic velocity. However, as $d(t)$ increases, the natural asymptotic steepness $\lambda^{**}(t)$ decreases and the front velocity eventually falls below the natural asymptotic velocity. There is no analogous phenomenon in the autonomous case, and na\"ive boundary conditions would not be able to capture this effect reliably or accurately because the shallow initial condition cannot be faithfully represented. We remark that, while a Robin boundary condition could in principle be imposed in the far-field, its usefulness is limited since the correct front steepness is only known once the solution is computed, and is, moreover, a function of time.


\section{Pushed fronts}\label{sec:pushed}

\subsection{Autonomous pushed fronts}
\begin{figure*}
\centering
\includegraphics[width=\textwidth]{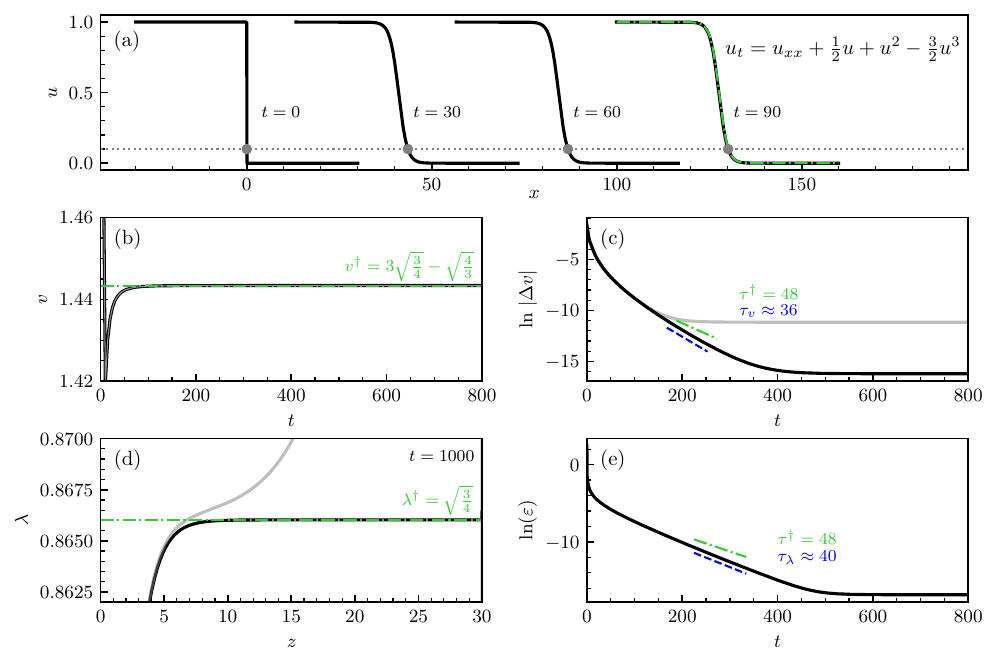}
\caption{The GBC method benchmarks well for the pushed front solutions of Eq.~(\ref{eq:fisher-23}) with $a=1/2$ and a step function initial condition. The simulation parameters are $dt=0.025$, $t\in[0,1\,000]$, $dz=0.001$, $z\in[-30, 30]$ in a frame moving with speed {$c = v^\dagger \equiv 3\sqrt{3/4} - \sqrt{4/3}$} and a buffer region of size $\delta = 10$. (a) The simulated front (solid, black) propagates through the lab frame and is visualized at four snapshots from $t=0$ to $t=90$. The simulated front agrees with the theoretical front solution (given in Eq.~(\ref{eq:pushed-theoretical-soln}), dash-dotted, green) at $t=90$. (b) The front velocity $v(t)$ in the lab frame compared between the GBC method (solid, black) and DZ method (solid, gray). Both methods approach the theoretically predicted front velocity (dash-dotted, green) and are visually indistinguishable. (c) Comparison of the velocity deviation $\Delta v \equiv v - v^{\dagger}$ between the two methods demonstrates a significant improvement in precision with the GBC method. The exponential convergence is in agreement with theoretical predictions \cite{rothe1981convergence}, although the extracted time constant via log-linear fitting, $\tau_v \approx 36$, deviates slightly from the expected $\tau^\dagger \approx 48$. The fitting region (dashed, blue) is chosen according to \cite{deshmukh2021toward} and compared against the expected time decay (dash-dotted, green). (d) The front steepness $\lambda = -u_z/u$ computed at $t=1\,000$ in the moving frame coordinate $z\equiv x - \int_0^t c(t')\,dt'$ using the GBC method tracks the predicted leading edge steepness $\lambda^{\dagger} = \sqrt{3/4}$ accurately while the DZ method yields an inaccurate value. (e) The deviation in front steepness is computed using the weighted norm metric, $\varepsilon$, for the GBC method over time. The time constant $\tau_\lambda \approx 40$ is extracted via log-linear fitting.} 
\label{fig:pushed_const}
\centering
\end{figure*}

To examine the pushed front regime, we consider the following modification to the F-KPP equation:
\begin{equation}{\label{eq:fisher-23}}
    u_{t} = u_{xx} + a u +  u^2 - (1+a)u^3.
\end{equation}
This quadratic-cubic nonlinearity is the same as that in the Nagumo equation \cite{NagumoEtAl1962,McKean1970}, except that in this case, following \cite{ebert_front_2000}, the parameter $a$ is now located in the linear and cubic terms. For $a > 1$, pulled fronts are selected in Eq.~(\ref{eq:fisher-23}). For $a<1$, pushed fronts emerge with asymptotic front steepness and velocity
\begin{align}
    \lambda^\dagger &= \sqrt{\frac{a + 1}{2}}, \label{eq:pushed-profile} \\ 
    v^\dagger &= 3 \lambda^\dagger - \frac{1}{\lambda^\dagger}. \label{eq:pushed-speed}
\end{align}
An analytical expression for the asymptotic front solution is also given in \cite{ebert_front_2000} (up to translation invariance):
\begin{equation}{\label{eq:pushed-theoretical-soln}}
u^\dagger(x,t) = \frac{e^{-\lambda^\dagger  (x-v^\dagger t)}}{1+e^{-\lambda^\dagger (x- v^\dagger t)}} \equiv u^\dagger(z) ,
\end{equation}
where the moving frame coordinate is $z=x-v^\dagger t$.

We simulate Eq.~(\ref{eq:fisher-23}) using the GBC method and find that it is in agreement with the theoretical predictions as depicted in Fig. \ref{fig:pushed_const}. The front profile approaches $u^\dagger$, as shown in \figsubref{fig:pushed_const}{a}. Since pushed front propagation is dictated by nonlinear bulk dynamics and the GBC method works exclusively on the leading edge, improved accuracy for pushed fronts is not necessarily expected. This is depicted in \figsubref{fig:pushed_const}{b}, where the velocity plots of the GBC and DZ methods nearly overlap. However, a closer look at the front velocity deviation in \figsubref{fig:pushed_const}{c} reveals a major difference in the precision of front tracking between the two methods. The GBC method improves agreement with the analytical solution by several orders of magnitude, indicating the importance of the appropriate boundary condition across regimes. 

The velocity of pushed fronts is known to converge exponentially \cite{rothe1981convergence}, and the rate of convergence can be predicted via a spectral stability analysis; see Appendix~\ref{app:lin_stab_pushed}. The theoretically predicted time constant of convergence is $\tau^{\dagger} = 48$, and we confirm this calculation using results from numerical simulations as depicted in \figsubref{fig:pushed_const}{c}. We perform a log-linear fit, where the fitting region is chosen according to \cite{deshmukh2021toward}, and numerically extract the velocity rate of convergence $\tau_v \approx 36$. Note that we are only able to perform fitting for early times after which the error $\Delta v\equiv v-v^\dagger$ tapers off. This is indicative of a persistent numerical error due to finite resolution and gives us a way to separate the discretization error from the boundary method error. The GBC method allows for a larger fitting region for the expected exponential convergence, thus providing a more reliable estimate of the velocity convergence time constant. 

As with pulled fronts, the GBC method also enables accurate measurements of the pushed front profile, particularly at the leading edge where na\"ive methods fail. We track the steepness $\lambda$ as depicted in \figsubref{fig:pushed_const}{d} and find agreement at late times with the asymptotic steepness $\lambda^{\dagger}$ in \eqref{eq:pushed-profile}. As with the velocity, the profile of pushed fronts also exhibits exponential convergence \cite{rothe1981convergence}, and we naturally expect the two rates to be the same. We numerically extract the profile convergence time constant $\tau_{\lambda}$ using a weighted energy functional metric, introduced in \cite{an2023quantitative} as 
\begin{equation}\label{eq:energy}
    \varepsilon(t) \equiv \sqrt{\int{e^{v^\dagger z}|U(z, t)|^2\,dz}},
\end{equation}
where $U(z,t)$ is the shape defect function defined as
\begin{equation}\label{eq:shape-defect}
    U(z,t) \equiv \zeta[u(z,t)] - u_z(z,t).
    \end{equation}
Here, $\zeta(u) \equiv -\lambda u(1-u)$ is the traveling wave profile function evaluated at the simulated field $u(z,t)$. The functional form of $\zeta$ is obtained by differentiating Eq.~\eqref{eq:pushed-theoretical-soln}, which gives $u_z^\dagger = -\lambda u^\dagger(1-u^\dagger)$; crucially, $\zeta$ is then evaluated at $u(z,t)$ rather than at $u^\dagger(z)$, which is what makes the resulting metric translationally invariant. By looking at the difference between the evolving and asymptotic fronts, we are able to account for the entire front profile in this metric. We follow the same fitting procedure as for $\tau_v$ to obtain $\tau_\lambda \approx 40$, as shown in \figsubref{fig:pushed_const}{e}. Note that the profile measurements, made possible by accurate leading edge data from the GBC method, are more robust than the velocity measurements. The latter rely on discrete time intervals in addition to discrete space; indeed, this higher degree of accuracy of the profile is consistent with the observed longer scaling region in \figsubref{fig:pushed_const}{e} compared to \figsubref{fig:pushed_const}{c}. Considering only $\tau_\lambda$, we conclude that this numerically measured value is reasonably close to the theoretically predicted $\tau^\dagger$. We attribute the discrepancy between $\tau_\lambda$ and $\tau^\dagger$ to a transient dominant mode. We do not attribute the discrepancy to any numerical simulation limitations as our result does not change with higher spatial resolution or larger domain size; see Fig.~\ref{fig:Ec_convergence} in Appendix~\ref{app:lin_stab_pushed}.

\subsection{Nonautonomous pushed fronts and the pushed--pulled transition}
\begin{figure}[hbt!]
\includegraphics[width=0.48\textwidth]{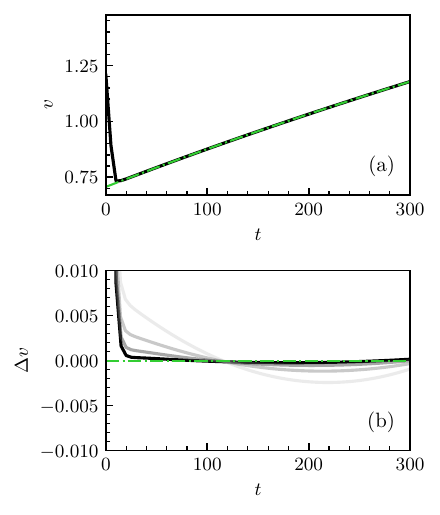}
\caption{A linearly increasing time-dependent parameter $a(t) = 0.001t$ demonstrates convergence to the time-frozen asymptotic front velocity (\ref{eq:pushed-time-frozen-speed}). The simulation parameters are $t \in [0,300]$, $dz = 0.001$, $z \in [-50, 200]$, $c=\gamma=0.9$ and $\delta=10$ with varying temporal resolutions. The time step size takes on the values $dt\in\{0.2, 0.1, 0.05, 0.025\}$ with smaller values depicted in progressively darker shades of gray in (b). These gray curves converge to the most accurate simulation at $dt=0.025$ (solid, black) indicating that as $dt \rightarrow 0$, the simulated front velocity converges to the time-frozen asymptotic velocity (dash-dotted, green).}
\label{fig:pushed_linear}
\end{figure}

We now modify Eq.~(\ref{eq:fisher-23}) to read
\begin{equation}\label{eq:pushed-time-dep}
    u_{t} = u_{xx} + a(t)u +  u^2 - [1+a(t)]u^3,
\end{equation}
starting with a linear time-dependence of the form 
\begin{equation}\label{eq:pushed-time-dep-param}
    a(t) = 0.001t
\end{equation}
for $t \in [0,300]$. Substituting Eq.~(\ref{eq:pushed-time-dep-param}) into Eq.~\eqref{eq:pushed-speed}, we obtain the time-frozen asymptotic speed as 
\begin{equation}\label{eq:pushed-time-frozen-speed}
    v^\dagger(t) = 3 \sqrt{\frac{0.001t + 1}{2}} - \frac{1}{0.001t}.
\end{equation}

We simulate Eq.~(\ref{eq:pushed-time-dep}) using the GBC method, the results of which are shown in Fig.~\ref{fig:pushed_linear}. We find that the numerically computed velocity quickly approaches the time-frozen asymptotic pushed speed \eqref{eq:pushed-time-frozen-speed} as the time step $dt$ is reduced. Since there is exponential convergence to the pushed front in the autonomous case, and we only impose linear growth for $a(t)$, this result seems reasonable.

\begin{figure*}
\centering
\includegraphics[width=\textwidth]{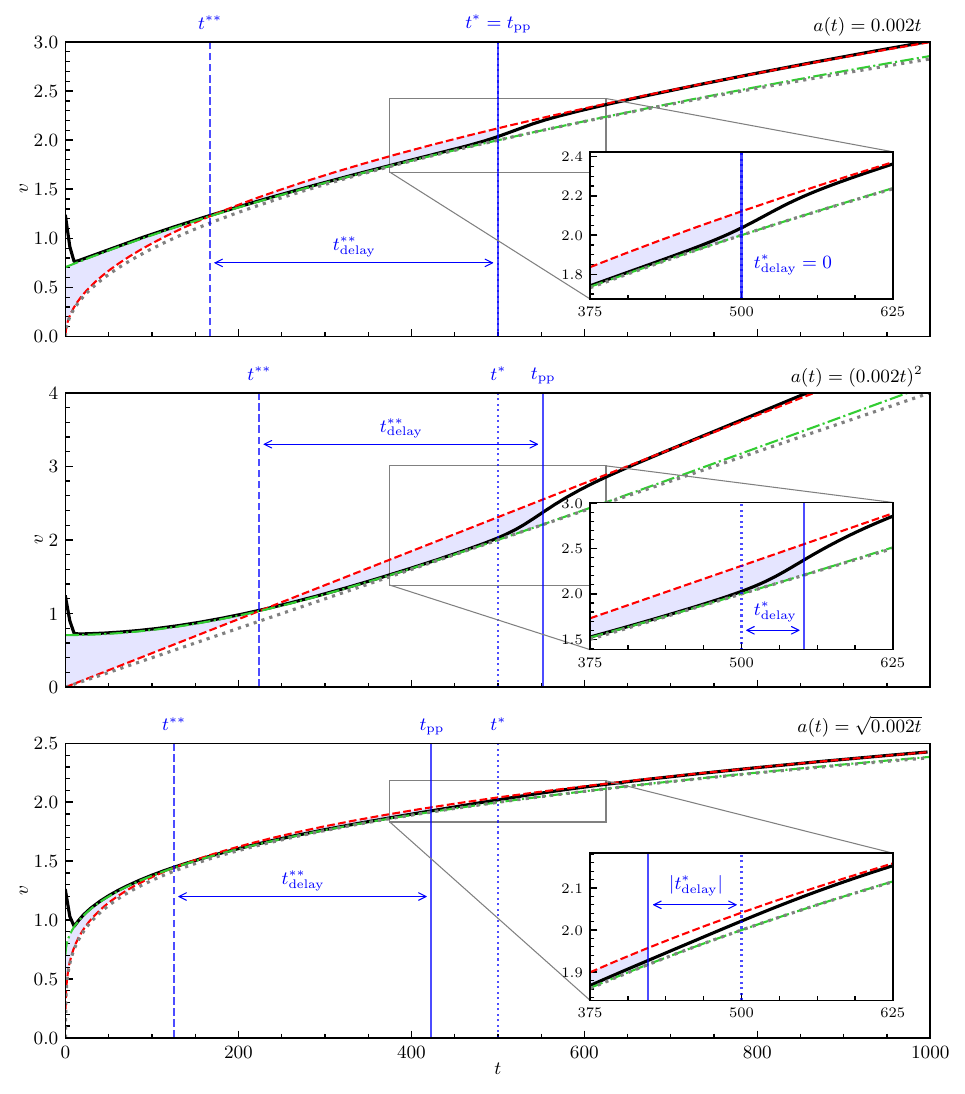}
\caption{Simulated pushed--pulled transitions in Eq.~\eqref{eq:pushed-time-dep} starting from a step initial condition for $a(t)=\epsilon t$ (top), $a(t)=(\epsilon t)^2$ (middle), and $a(t)=\sqrt{\epsilon t}$ (bottom) with $\epsilon = 0.002$ show zero, positive, and negative bifurcation delay, respectively, in agreement with theory. Initially, the measured velocity (solid, black) follows the time-frozen pushed velocity $v^\dagger(t)$ (dash-dotted, green). Then, the front leaves the pushed regime and approaches the natural pulled asymptotic velocity $v^{**}(t)$ (dashed, red). The time-frozen pulled asymptotic velocity $v^*(t)$ (dotted, gray) is shown for reference. In each case, we observe that the pushed--pulled transition occurs near $t_{\textrm{pp}}$ (solid, blue), the theoretical transition point \eqref{eq:t_pp} at which the integral of the velocity difference $v^{**}(t) - v^{\dagger}(t)$ (shaded blue region) vanishes. The velocity crossing times $t^{*}$ [\eqref{eq:t_star}, dotted blue] and $t^{**}$ [\eqref{eq:t_starstar}, dashed blue] and the corresponding delays $t_{\mathrm{delay}}^*$ \eqref{eq:t_delay_star} and $t_{\mathrm{delay}}^{**}$ \eqref{eq:t_delay_starstar} are also shown.}
\label{fig:transition}
\centering
\end{figure*}

Given the above analysis, pushed front convergence appears straightforward, but below we reveal more subtle behavior by analyzing the dynamic pushed-to-pulled transition. This transition is an ideal phenomenon to study with the GBC method for its inherent time-dependence and long-time pulled behavior.

Recall that in the autonomous case of Eq.~\eqref{eq:fisher-23}, pushed fronts are selected for $a<1$ and pulled fronts are selected for $a>1$. At the pushed--pulled transition $a=1$, the pushed and pulled fronts converge along with their asymptotic velocities. The pushed--pulled transition is a codimension-one bifurcation \cite{avery_pushedpulled_2022} and is an example of a global bifurcation: it is associated with a change in the qualitative behavior of a heteroclinic orbit in the phase plane \cite{van_saarloos_front_2003}.

The dynamic crossing of a bifurcation point can lead to a phenomenon known as \textit{bifurcation delay} \cite{benoit_dynamic_1991,tsubota_bifurcation_2024}, in which the onset of the bifurcation occurs later than the time at which the bifurcation parameter crosses the time-frozen critical value. Delay phenomena in pushed--pulled transitions have recently been observed \cite{AVERY2025134972}.

Bifurcation delay arises in our problem via pushed--pulled competition. For concreteness, we take $a(t)$ to be algebraically increasing, e.g. $a(t)=(\epsilon t)^{\alpha}$ for $\epsilon, \, \alpha > 0$. Initially, the pushed front velocity is faster than the pulled front velocity, so in the comoving frame of the pushed front, the linear equation is only convectively unstable; that is, linear perturbations are left behind the comoving frame \cite{van_saarloos_front_2003}. However, as $t$ increases, the linear equation becomes absolutely unstable, and we expect the linear modes to catch up and take over.

When does the convective instability become an absolute instability in the frame of the pushed front? It is natural to assume that this turnaround will occur when $a(t)=1$, but this is not the case. For algebraically increasing $a(t)$, the natural asymptotic velocity $v^{**}(t)$ \eqref{eq:natural_asymptotic_velocity} is larger than the time-frozen one $v^{*}(t)$ \eqref{eq:time_frozen_asymptotic_velocity}. Therefore, $v^{**}(t)$ attains (and passes) $v^\dagger(t)$  at an earlier time than when $a(t)=1$, so the turnaround occurs earlier. However, at the turnaround, the linear modes have receded; it is only when they catch up that we see the result: a pushed--pulled transition.

With this idea in mind, we can define the \textit{turnaround times} $t^*$ and $t^{**}$ such that
\begin{subequations}
\begin{align}
    v^{\dagger}(t^*) &= v^*(t^*), \label{eq:t_star} \\
    v^{\dagger}(t^{**}) &= v^{**}(t^{**}). \label{eq:t_starstar}
\end{align}
\end{subequations}
Then, we can define the theoretical \textit{pushed--pulled transition time} $t_{\mathrm{pp}} > 0$ as the time that solves
\begin{equation}\label{eq:t_pp}
    \int_0^{t_{\mathrm{pp}}} [v^{**}(t) - v^\dagger(t)]\,dt = 0.
\end{equation}
This articulates the discussion above. For $t < t_{\mathrm{pp}}$, the pushed front moves faster. For $t > t_{\mathrm{pp}}$, the pulled front moves faster. For algebraic growth, the integral \eqref{eq:t_pp} has a closed form; see Appendix~\ref{app:pushed_pulled_transition_time}. Finally, we can define delay times
\begin{subequations}
\begin{align}
    t_{\mathrm{delay}}^* &\equiv t_{\mathrm{pp}} - t^*, \label{eq:t_delay_star} \\
    t_{\mathrm{delay}}^{**} &\equiv t_{\mathrm{pp}} - t^{**} .\label{eq:t_delay_starstar}
\end{align}
\end{subequations}
We interpret $t_{\mathrm{delay}}^*$ as bifurcation delay; it is the additional time after the time-frozen bifurcation point before observing the effect of the bifurcation. We interpret $t_{\mathrm{delay}}^{**}$ as the time it takes for the linear modes to take over after the turnaround. We do not refer to this quantity as bifurcation delay, since it does not refer to a time-frozen bifurcation -- it is a distinct quantity.

In Fig.~\ref{fig:transition}, we simulate the dynamic crossing of the pushed--pulled transition for linear, quadratic, and square root growth of $a(t)$. The data confirms the idea outlined above: the pushed--pulled transition occurs when Eq.~\eqref{eq:t_pp} holds. For linear growth, coincidentally $t_{\mathrm{pp}}=t^*$, thus $t_{\mathrm{delay}}^*=0$ and no bifurcation delay is observed (see Appendix~\ref{app:pushed_pulled_transition_time}). This behavior is not generic across other forms of algebraic growth. For quadratic growth, we observe bifurcation delay $t_{\mathrm{delay}}^*>0$. For square root growth, we remarkably find $t_{\mathrm{delay}}^*<0$, i.e. \textit{negative bifurcation delay}. In each case, the ordering of $t_{\mathrm{pp}}$ and $t^*$ is independent of the growth rate $\epsilon$. Of course, in each case, we also observe $t_{\rm delay}^{**} > 0$.

\begin{figure}
\includegraphics[width=0.48\textwidth]{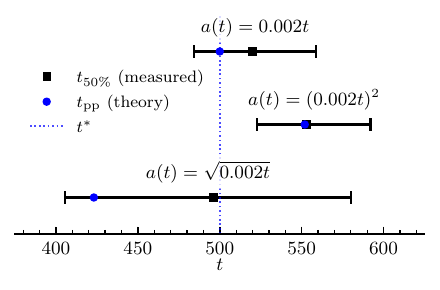}
\caption{Timing of the simulated pushed--pulled transitions in Fig.~\ref{fig:transition} confirms theoretical predictions. For each simulation, we measure the times at which the transition progress \eqref{eq:transition_progress}, defined in terms of the velocity difference from $v^{\dagger}(t)$, is 20\% (left endpoint), 50\% (middle black square), and 80\% (right endpoint). The theoretical transition time $t_{\mathrm{pp}}$ (solid blue) and the time-frozen transition time $t^*$ (dashed blue) are drawn for reference. In each case, we find that $t_{\mathrm{pp}}$ accurately describes the transition time. The 20\% and 50\% completion times are ordered correctly. Notably, for square root growth, the transition is more than halfway complete by $t^*=500$, thereby confirming a negative bifurcation delay. The transition durations vary, and the 80\% point deviates from the predicted ordering, but these facts are not surprising owing to the different asymptotic behaviors of each of the pulled regimes.}
\label{fig:transition_timing}
\end{figure}

To compare the simulations with theory more precisely, we define an empirical measure of the transition progress as the relative velocity deviation from $v^{\dagger}(t)$:
\begin{equation}\label{eq:transition_progress}
    \frac{v(t) - v^{\dagger}(t)}{v^{**}(t) - v^{\dagger}(t)} \times 100\%.
\end{equation}
In Fig.~\ref{fig:transition_timing}, we measure the 20\%, 50\%, and 80\% completion times. The 20\% completion time characterizes the \textit{exit time} when the solution leaves the neighborhood of the original attractor; neighborhood exit is the typical definition of bifurcation onset in the context of bifurcation delay (cf. \cite{tsubota_bifurcation_2024}). In all cases, the measured 20\% and 50\% completion times in Fig.~\ref{fig:transition_timing} are qualitatively consistent with the transition time predicted by $t_{\textrm{pp}}$. The 80\% completion time characterizes the \textit{entrance time} into the neighborhood of the new attractor, and this time depends on the asymptotic properties of the attractor. Measuring the entrance time implicitly assumes that the front speed converges to the natural asymptotic velocity. This is not true in general (Fig.~\ref{fig:d_linear}) but is expected for monotonically increasing $a(t)$ (see Sec.~\ref{sec:pulled}). Further, we speculate that algebraic growth of $a(t)$ with a larger power of $t$ causes faster convergence to $v^{**}(t)$. This is consistent with Fig.~\ref{fig:transition_timing}, where the transition duration is indeed seen to shorten  with increasing power.

We conclude this section with a few additional remarks. Although our results are consistent, we recognize that it is not clear how to precisely define the onset of the transition. The theoretical value $t_{\mathrm{pp}}$ as defined by Eq.~\eqref{eq:t_pp} does not consider transient behavior that would add additional terms depending on the initial condition. Additionally, the empirical measure proposed in Eq.~\eqref{eq:transition_progress}, defined via velocities, is likely inadequate. In fact, the height $H$ at which the front is tracked results in slightly different velocity curves (see Eq.~\eqref{eq:linear_front_position_exact}), although for the convergent cases we expect $H$ dependence to appear only in higher-order terms as is true in the autonomous case \cite{van_saarloos_front_2003}. A proper definition should rely on an underlying metric to characterize the distance between the asymptotic pushed and pulled fronts. One can then precisely define the neighborhoods to determine the exit time and transition duration \cite{tsubota_bifurcation_2024}.

Finally, in the original example of a nonautonomous pushed front (Fig.~\ref{fig:pushed_linear}), the simulation time is restricted so that $v^{**}(t) < v^{\dagger}(t)$ and thus $t<t^{**}$ for all $t$. Therefore, the simulation remains within the pushed regime. However, in light of these results, it is important to consider that, in general, pulled front competition must be considered even in cases that appear entirely pushed if analyzed adiabatically.

\section{Discussion}\label{sec:discussion}

The GBC method enables accurate simulation of front propagation in the F-KPP equation by enforcing a time-dependent boundary condition at the leading edge, enabling precise measurements of front velocities even on relatively small computational domains. The GBC benchmarks well against known results for the autonomous Fisher equation with both steep and shallow initial conditions. This contrasts with previous approaches, which typically require different boundary conditions for steep and shallow initial data. Furthermore, the GBC method can handle time-dependent growth and diffusion coefficients, revealing complex nonlinear behavior that deviates from the predictions of linear theory for pulled fronts. The same numerical framework is also capable of precisely simulating pushed fronts and the pushed--pulled transition, revealing both delayed and premature transition onset dictated by the competition between the two regimes for different forms of time-depedent coefficients.


Much remains to be understood about the nature of fronts with a time-dependent parameter. How the natural asymptotic velocity and profile characterize pulled fronts beyond the simplest cases remains poorly understood. A more careful approach is needed to understand the asymptotics of pulled fronts beyond the simple linear analysis presented here. The influence of time-dependent parameters on shallow initial conditions is also largely unexplored. The dynamic transition from the ``shallow'' to the ``steep'' regime and vice versa is not a phenomenon that can occur in the autonomous problem. For pushed fronts, the pushed–pulled transition offers a particularly rich setting for exploring global bifurcation delay in a spatially extended system. The delayed and premature onset of the transition joins the set of novel phenomena induced by a time-dependent forcing along with other nonautonomous behaviors such as rate-induced tipping \cite{WieczorekEtAl2023}. Understanding the geometry and topology of the pushed--pulled transition would be valuable for improving the robustness of the current results. We now have a powerful numerical tool to explore these questions and more.


A rigorous numerical analysis of the GBC method, beyond the proof-of-concept validation presented here, remains an important direction for future work. Perhaps most importantly, we need a better understanding of the accuracy and stability of this method. We also have a limited understanding of how to select the buffer size $\delta$ for general time-dependent parameters. We also restrict ourselves here to a Dirichlet-to-Dirichlet boundary condition, where the NL region provides a Dirichlet condition for the LA region, which in turn provides a Dirichlet condition for the NL region. One could consider other combinations, e.g. Dirichlet-to-Neumann or Neumann-to-Dirichlet, which may have different stability and accuracy properties. Finally, we have not considered the derivable corrections to the front velocity due to discretization \cite{van_saarloos_front_2003,ebert_front_2000,ponedel_front_2017}.

The foundational finite-difference scheme can also be improved. Here, we used a simple first-order IMEX Euler scheme in time with second-order central finite differences in space. Parabolic equations such as F-KPP are often simulated with Crank-Nicolson, other IMEX/implicit schemes, or other specialized schemes \cite{hagstrom_numerical_1986,ebert_front_2000, mickens_best_1994}. To resolve the time-dependent parameters more accurately, one may consider a smaller time step. However, this would require optimizations or approximations to the GBC method to avoid excessive numerical integration time to compute the boundary value (see Appendix~\ref{app:numerical}).

Finally, we can consider more complex equations. The GBC method can currently only be applied to the one-dimensional nonautonomous F-KPP equation with a somewhat restrictive requirement on the moving frame speed. These restrictions arise from the need to construct a half-line Green’s function and evaluate the linear solution in closed form via Gaussian integrals. Developing analytical and/or numerical tools to relax these restrictions---by approximating the relevant integrals, for instance---would enable the GBC method to be applied to a wider variety of problems, including those with higher-order spatial derivatives or even spatial heterogeneities \cite{kuske_pattern_1997, GohEtAl2023}.


\begin{acknowledgments}
We gratefully acknowledge helpful comments and suggestions from C. Liu and V. Klika. This work was supported by the Berkeley Physics-and-Astronomy Undergraduate Research Scholars (BPURS) Program (T.T. and S.M.), a US Dynamics Days travel stipend (S.M.) and by the National Science Foundation under grant DMS-2308337 (T.T. and E.K.).
\end{acknowledgments}

\section*{Code Availability}
The code used to perform the simulations is publicly available at \url{https://github.com/tktsubota/fronts-gbc}.

\bibliography{references.bib}

\onecolumngrid
\appendix

\section{Theoretical Details}\label{app:theoretical}

We seek an exact solution of the linearized, nonautonomous F-KPP equation on the half-line $z\in[0,\infty)$ with an arbitrary initial condition and an arbitrary time-dependent Dirichlet boundary value:
\begin{subequations}\label{eq:linearized_general_all}
\begin{align}
    u_t &= a(t)u + c(t)u_z + d(t)u_{z}, \label{eq:linearized_general} \\
    u(z,0) &= f(z), \\
    u(0,t) &= g(t).
\end{align}
\end{subequations}
We begin by finding the free Green's function, i.e. the Green's function for the infinite line. By taking the Fourier transform of \eqref{eq:linearized_general} in $z$, we obtain 
\begin{equation}
   \hat{u}_t(k,t) = \left\lbrace -kc(t) + i[a(t) - k^2d(t)]\right\rbrace  \hat{u}(k,t).
\end{equation}
The resulting free Green's function for delta forcing at $(y,s)$ is
\begin{equation}
    G_{\mathrm{free}}(z,t;y,s) = \frac{1}{\sqrt{4\pi \Tilde{D}(t,s)}}\exp\left[\Tilde{A}(t,s) - \frac{(z-y+\Tilde{C}(t,s))^2}{4\Tilde{D}(t,s)}\right],
\end{equation}
where $\Tilde{A}(t,s) = A(t) - A(s)$, with $A(t)=\int_0^t a(t')dt'$ as in Eq.~\eqref{eq:natural_asymptotic_velocity}, and $\Tilde{C}(t,s)$ and $\Tilde{D}(t,s)$ defined similarly.

We can now use the free Green's function to construct the half-line Green's function. In the case $g(t) = 0$, we use the method of images. Na\"ively, we may expect to use a symmetric image charge, but advection due to the moving frame breaks the reflection symmetry. To resolve this matter, we take inspiration from \cite{redner_guide_2001}, seeking a weighted image charge that satisfies the desired boundary condition $u(0,t) = 0$:
\begin{equation}
    0 = G_{\mathrm{free}}(0,t;y,s) - \alpha G_{\mathrm{free}}(0,t;-y,s). \label{eq:image_weighted}
\end{equation}
We solve Eq.~\eqref{eq:image_weighted} to find that
\begin{equation}
    \alpha = \exp\left[\frac{c(t)}{d(t)}y\right].
\end{equation}
In order for Eq.~\eqref{eq:image_weighted} to solve Eq.~\eqref{eq:linearized_general_all}, the weight $\alpha$ must be time-independent. Hence, to proceed further, we require that $c(t)$ be proportional to $d(t)$. We define the proportionality constant $\gamma$ so that $c(t) = \gamma d(t)$ and $\alpha =e^{\gamma y}$. The half-line Green's function is
\begin{equation}
    G(z,t;y,s) = G_{\mathrm{free}}(z,t;y,s) - e^{\gamma y}G_{\mathrm{free}}(z,t;-y,s)
\end{equation}
and the solution to the original problem, with $g(t)=0$, is then
\begin{equation}
    u(z,t) = \int_0^{\infty} G(z,t;y,0)f(y)\,dy.
\end{equation}

For the case $g(t)\neq 0$, we may na\"ively proceed with the change-of-variable $v(z,t) = u(z,t) - g(t)$. This substitution converts the boundary condition into an inhomogeneous forcing $F(t) = -g'(t)$. However, because this forcing is constant for all $x$ and $\lim_{t\to\infty} \int_0^\infty G(z,t;y,s) \,dy$ is not finite, the resulting solution is not well-behaved as $t\to\infty$. Therefore, we pick a different substitution that enforces regularity. The inhomogeneous forcing must decay as $x\to\infty$, so we consider the substitution
\begin{equation}
    w(z,t) = u(z,t) - g(t)e^{-z^2}.
\end{equation}
Note that an exponential decay term $g(t)e^{-rz}$ would require tuning the decay rate $r$ to be large enough to ensure a finite limit; the Gaussian term is thus a preferable substitution. The resulting equation is
\begin{subequations}
\begin{align}
    w_t &= a(t)w + c(t)w_z + d(t)w_{zz} + F(z,t) \\
    w(z,0) &= f(z) - g(0)e^{-z^2} \\
    w(0,t) &= 0 \\
    F(z,t) &= [(a(t) - 2d(t))g(t) - g'(t)]e^{-z^2} - 2c(t)g(t)ze^{-z^2} + 4d(t)g(t)z^2e^{-z^2}.
\end{align}
\end{subequations}
By appealing to Duhamel's principle \cite{evans_partial_2010}, we can write down the exact solution to Eq.~\eqref{eq:linearized_general_all} as
\begin{equation}\label{eq:greens_exact}
    u(z,t) = g(t)e^{-z^2} + \int_0^{\infty} G(z,t;y,0)[f(y)-g(0)e^{-y^2}]\,dy + \int_0^t \int_0^\infty G(z,t;y,s)F(y,s)\,dy\,ds.
\end{equation}
The final term may be written as 
\begin{equation}\label{eq:greens_integrals}
    \int_0^t \int_0^\infty G(z,t;y,s)F(y,s)\,dy\,ds = \int_0^t [h_0(s) I_0(z,t;s) + h_1(s) I_1(z,t;s) + h_2(s) I_2(z,t;s)]\,ds,
\end{equation}
where
\begin{subequations}
\begin{align}
    h_0(t) = [a(t) - 2d(t)]g(t) - g'(t),& \ \  I_0(z,t;s) = \int_0^\infty G(z,t;y,s)e^{-y^2}\,dy, \label{eq:h0I0} \\
    h_1(t) = - 2c(t)g(t),& \ \ I_1(z,t;s) = \int_0^\infty G(z,t;y,s)ye^{-y^2}\,dy, \\
    h_2(t) = 4d(t)g(t),& \ \ I_2(z,t;s) = \int_0^\infty G(z,t;y,s)y^2e^{-y^2}\,dy.
\end{align}
\end{subequations}
Since $G(z,t;y,s)$ is of Gaussian form, the integrals $I_k$ have closed-form expressions which all converge to a finite limit as $t\to\infty$.

\section{Numerical Details}\label{app:numerical}

\subsection{Detailed algorithm}

The numerical algorithm for the GBC method is built upon a standard finite-difference method. Formally, let $N_z$ and $N_t$ be the number of spatial grid points (in the moving frame coordinate $z$) and time steps, respectively. For the field $u(z,t)$, define the discrete numerical approximation as $\tilde{u}(z_j,t^{(i)}) = \tilde{u}_j^{(i)}$, where $t^{(i)}=i \, dt$ for $i=0,\dots,N_t-1$ and $z_j=z_0 + j \, dz$ for $j=0,\dots,N_z$. Additionally, define $a(t^{(i)})=a^{(i)}$, $c(t^{(i)})=c^{(i)}$, and $d(t^{(i)})=d^{(i)}$.

The finite difference equation at time $t^{(i)}$ for $j = 1,\dots,N_z-1$, with second-order central differences in space and an implicit-explicit (IMEX) Euler scheme in time, is 
\begin{subequations}\label{eq:finite_difference}
\begin{equation}
    \frac{\tilde{u}^{(i+1)}_j - \tilde{u}^{(i)}_j}{dt} = a^{(i+1)}\tilde{u}^{(i+1)}_j + c^{(i+1)}\left[\frac{\tilde{u}^{(i+1)}_{j+1}-\tilde{u}^{(i+1)}_{j-1}}{2\,dz}\right] + d^{(i+1)}\left[\frac{-\tilde{u}^{(i+1)}_{j+1}+2\tilde{u}^{(i+1)}_j-\tilde{u}^{(i+1)}_{j-1}}{(dz)^2}\right] + \mathcal{N}\left(\tilde{u}^{(i)}_j,t^{(i)}\right),
\end{equation}
where $\mathcal{N}(u,t)$ are the nonlinear terms in $f(u,t)$ which are treated explicitly. For the boundary terms, we have
\begin{align}
    \tilde{u}_0^{(i+1)} = \tilde{u}_1^{(i+1)} \\
    \tilde{u}_{N_z}^{(i+1)} = \tilde{b}^{(i+1)}.
\end{align}
\end{subequations}
Here, $\tilde{b}^{(i)} = \tilde{b}(t^{(i)})$ is the GBC Dirichlet boundary value. The result constitutes a system of $N_z+1$ linear equations, which can be solved to determine $\tilde{u}^{(i+1)}$.

As depicted in Fig.~\ref{fig:numerical_approach}, we compute $\tilde{b}^{(i+1)}$ prior to the implicit step. To do so, we build an array of $\tilde{g}^{(i)} \equiv \tilde{u}\left(z_1-\delta,t^{(i)}\right)$ and its time derivatives $\tilde{g}'^{(i)}$. At time $t^{(i)}$, we perform the following steps:
\begin{enumerate}
    \item Measure the current value $\tilde{g}^{(i)}$.
    \item Compute $\tilde{g}'^{(i)}$ as a first-order right-handed derivative.
    \item Compute the integral \eqref{eq:greens_integrals} using the trapezoidal rule with step size $dt$.
    \item Compute the final result \eqref{eq:greens_exact}. This is $\tilde{b}^{(i+1)}$.
\end{enumerate}
Because the integrals $I_k$ are independent of the simulation, we can precompute the $N_t^2$ values for each $I_k$. We can also precompute the $N_t$ values for the initial condition integral (the second term in Eq.~\eqref{eq:greens_exact}). Thus, the dominant compute time contribution in computing $\tilde{b}^{(i+1)}$ is  Eq.~\eqref{eq:greens_integrals}, which takes $\ord(N_t)$ time steps. Solving Eq.~\eqref{eq:finite_difference} using a sparse solver takes $\ord(N_z)$. Altogether, the whole algorithm takes $\ord(N_t^2 + N_t N_z)$ time.


\subsection{Convergence}

In the main text, we have provided numerical evidence indicating that the method converges as $dt\to 0$. In particular, in Figs.~\ref{fig:constant_steep}, \ref{fig:constant_shallow}, and \ref{fig:pushed_const}, we demonstrate good agreement with existing theoretical predictions of the front speed in the autonomous case. We also have agreement, in Fig.~\ref{fig:a_linear}, with recent results on convergence for linear growth in the growth parameter \cite{AVERY2025134972}.

\begin{figure}
\includegraphics[width=\textwidth]{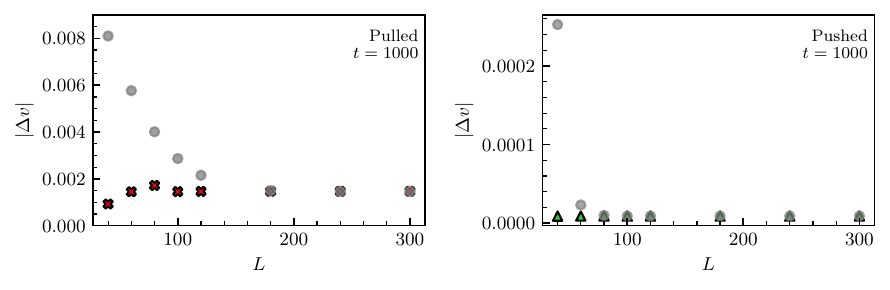}
\caption{For the autonomous F-KPP equation (see Figs.~\ref{fig:constant_steep} and \ref{fig:pushed_const}) in computational domains of large size $L=x_1-x_0$, the GBC method measures the same long-time front velocity as the DZ method (gray dots) in both the pulled (red X's) and pushed (green triangles) regimes. For small domain sizes $L$, the GBC method outperforms the DZ method, with larger improvements for pulled fronts, as expected. Here, $\Delta v$ is the deviation from the relevant asymptotic velocity at the finite time $t=1\,000$, so this does not necessarily capture error from ground truth. Simulation parameters are $dt=0.025$, $dz=0.01$, and $\delta=10$.}
\label{fig:L_convergence}
\end{figure}

In Fig.~\ref{fig:L_convergence}, we show furthermore that the GBC method converges to the DZ method as the domain size $L\to\infty$, as expected. The convergence in the pushed regime is much faster than in the pulled regime, as expected generically for numerical methods for fronts \cite{avery2025selection}. Importantly, the GBC method faithfully returns the correct front speed even with relatively small $L$, thus eliminating the need for a large domain size.

\begin{figure}
\includegraphics[width=0.48\textwidth]{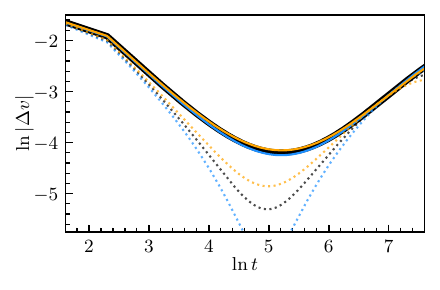}
\caption{The choice of $\gamma=c(t)/d(t)$ affects the numerically measured velocity deviation $\Delta v$. For the nonautonomous pulled front with $d(t) = 1 + \epsilon t$, $\epsilon=0.001$, the original choice of $\gamma=1.40$ (black), is reproduced from \figsubref{fig:d_linear}{c}, while alternative choices $\gamma = 1.35$ (blue) and $\gamma=1.45$ (orange) are overlaid. A comparison of the curves for $dt=0.025$ (solid) with $dt=0.2$ (dotted) shows that a different choice of $\gamma$ becomes less important as $dt$ is reduced.}
\label{fig:gamma_selection}
\end{figure}

Our last convergence result concerns the frame speed, i.e. the choice of $\gamma = c(t)/d(t)$. Ideally, one would select $\gamma$ so that the front is near-stationary in the moving frame. Roughly speaking, if the solution varies slowly relative to the time step, the error is smaller. More precisely, the error for Euler's method (explicit and implicit) applied to a one-dimensional ordinary differential equation is bounded by $dt \cdot \max |y''(t)|$, where $y''(t)$ is the second derivative of the solution \cite{BurdenFaires2011}. Therefore, for the constant parameter case, we select the asymptotic speed for $\gamma$. However, for a nonautonomous regime, there is no natural choice of $\gamma$ because the front shifts relative to the moving frame; this may induce a larger error. In Fig.~\ref{fig:gamma_selection}, we do find differences between $\gamma=1.35,1.40$, and $1.45$ for the nonautonomous pulled fronts we simulated in Fig.~\ref{fig:d_linear}. However, as expected, as $dt\to 0$, the choice of $\gamma$ becomes unimportant for the main conclusions.

\subsection{Buffer size selection}

\begin{figure}
\includegraphics[width=0.48\textwidth]{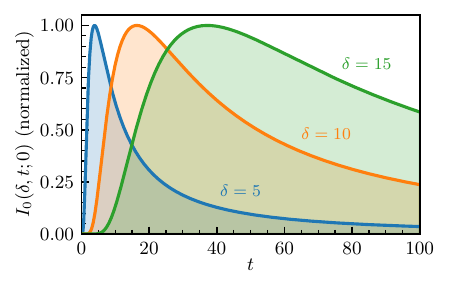}
\caption{Effect of buffer size $\delta$ on the normalized integration kernel $I_0(x=\delta,t;s=0)$, Eq.~\eqref{eq:h0I0}, in the autonomous case $a(t)=d(t)=1$ and $c(t)=\gamma=2$. Three buffer sizes $\delta=5,10$, and $15$ are shown in blue, orange, and green, respectively, indicating that a smaller buffer size results in a narrower, sharper integration kernel.}
\label{fig:buffer_size}
\end{figure}

Figure~\ref{fig:buffer_size} shows the kernel, defined in Appendix~\ref{app:theoretical}, $I_0(z=\delta,t;s=0)$ for $\delta=5, 10$ and $15$ in the constant Fisher regime described in Fig.~\ref{fig:constant_steep}, i.e. $a(t)=d(t)=1$ and $c(t)=\gamma=2$. We see that a small buffer region results in a narrow, sharp integration kernel, while a large buffer region results in a wide, flat kernel. Thus, in general, for a fixed $dt$, a large buffer region minimizes the numerical integration error. However, an excessively large buffer region would only add unnecessary computation time due to the larger domain.

Additionally, upon examining Fig.~\ref{fig:buffer_size}, it is natural to consider a cutoff for the integral \eqref{eq:greens_integrals} by setting $G=0$ for $t-s > \Delta$. This would reduce the numerical integration time from $\ord(N_t)$ to $\ord(1)$ and reduce the initial overhead of precomputing each $I_k$ from $\ord(N_t^2)$ to $\ord(N_t)$. This cutoff window $\Delta$ would be smaller for smaller choices of $\delta$. Because we sought exact expressions (Appendix~\ref{app:theoretical}) in our study, this is not a direction we pursue here but it is a worthwhile optimization to consider.

\section{Linear Stability Analysis of Pushed Fronts \label{app:lin_stab_pushed}}
Here we analyze the linear stability of the pushed front profile given in Eq.~(\ref{eq:pushed-profile}), $u=u^\dagger(z)$ with  $z=x-v^\dagger t$, which solves Eq.~(\ref{eq:fisher-23}). We write $u(x,t)=u^\dagger(z) + \eta(z,t)$ and linearize in the perturbation $\eta$ to obtain
\begin{equation}
    \label{eq:perturbation_eqn_about_u_dagger}
    \eta_t=\eta_{zz} + v^\dagger \eta_z + f'(u^\dagger) \eta,
\end{equation}
where $f'(u^\dagger)= a + 2u^\dagger - 3(1+a)(u^\dagger)^2$. We introduce $\eta = e^{\sigma t} \psi(z)$, where $\sigma$ is the growth rate, such that
\begin{equation}
    \sigma \psi = \psi'' + v^\dagger \psi' + f'(u^\dagger) \psi.
\end{equation}
We eliminate the first-order term using the substitution $\psi(z) \equiv e^{-v^\dagger z/2} \phi(z)$, giving
\begin{equation}
\sigma \phi = \phi '' - \frac{(v^\dagger)^2}{4}\phi+f'(u^\dagger)\phi
\end{equation}
which is the stationary Schrödinger equation
\begin{equation}
    -\phi'' + V(z) \phi \equiv \mathcal{L}\phi =E {\phi}, \label{eq:schrodinger}
\end{equation}
with energy $E=-\sigma$ and the nontrivial potential
\begin{equation}
V(z)
=
\frac{(v^\dagger)^2}{4}
-a
-2u^\dagger(z)
+
3(1+a)(u^\dagger(z))^2.
\end{equation}
Substituting the exact profile for $u^\dagger(z)$ from Eq.~(\ref{eq:pushed-theoretical-soln}) gives
\begin{equation}
V(z)
=
\frac{(v^\dagger)^2}{4}
- a
-\frac{2}{1+e^{\lambda^\dagger z}}
+
\frac{3(1+a)}
{\left(1+e^{\lambda^\dagger z}\right)^2}. \label{eq:potential_lin_stab_pushed}
\end{equation}

The potential $V(z)$ asymptotes to the constant $V_+\equiv(v^\dagger)^2/4 - a$ ahead of the front ($z\to \infty$) and $V(z)\to V_-\equiv (v^\dagger)^2/4+2a+1$ behind the front ($z\to -\infty$). The corresponding operators
\begin{equation}
    \mathcal{L}_\pm = -\frac{\mathrm{d}^2}{\mathrm{d}z^2} + V_\pm\label{eq:schrodinger_operators}
\end{equation}
are constant-coefficient Schrödinger operators whose spectra are given by
\begin{equation}
\sigma(\mathcal{L}_\pm)
=
\left\{V_\pm+k^2:\;k\in\mathbb{R}\right\}
=
[V_\pm,\infty).
\end{equation}

Since \(V(z)-V_\pm\) decays exponentially as \(z\to\pm\infty\), the operator $\mathcal{L}
\equiv
-\frac{\mathrm{d}^2}{\mathrm{d}z^2}
+V(z)$
represents a localized perturbation of the asymptotic operators $\mathcal{L}_\pm$. By Weyl's theorem \cite[Theorem~14.6]{HislopSigal} (see also \cite{cycon1987schrodinger}), localized perturbations do not modify the essential spectrum. Hence

\begin{equation}
\sigma_{\mathrm{ess}}(\mathcal{L})
=
\sigma(\mathcal{L}_+)\cup\sigma(\mathcal{L}_-)
=
[\min(V_+,V_-),\infty)
=
[V_+,\infty),
\end{equation}
since \(V_+<V_-\) for all \(a>0\).

The discrete spectrum contains the eigenvalue $0$ due to the translation invariance of the problem. To verify whether there are any discrete eigenvalues between $0$ and $V_+$, we computed the discrete spectrum using a finite-difference method on a large truncated interval with Dirichlet zero boundary conditions (not shown). No discrete eigenvalue greater than $0$ and less than $V_+$ was observed. We therefore conclude that, according to our linear stability analysis, $V_+$ is the asymptotic rate of convergence to the pushed front. For the example $a=1/2$, this indicates that the time scale of convergence is expected to be $\tau=1/V_+=48$.


\begin{figure}
    \centering
    \includegraphics[width=0.48\textwidth]{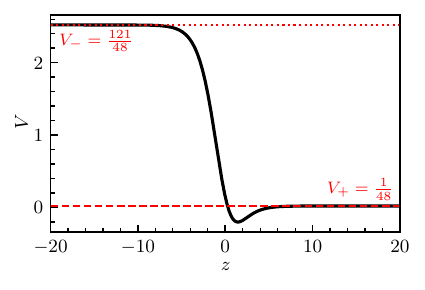}
    \caption{Potential $V(z)$ (solid, black) defined in Eq.~(\ref{eq:potential_lin_stab_pushed}) for $a=1/2$. The potential asymptotes to $V_+=1/48$ (dashed, red) at the leading edge, and to $V_-=121/48$ (dotted, red) at the trailing edge.}
    \label{fig:potential_lin_stab_pushed_front}
\end{figure}

\begin{figure}
    \centering
    \includegraphics[width=1\linewidth]{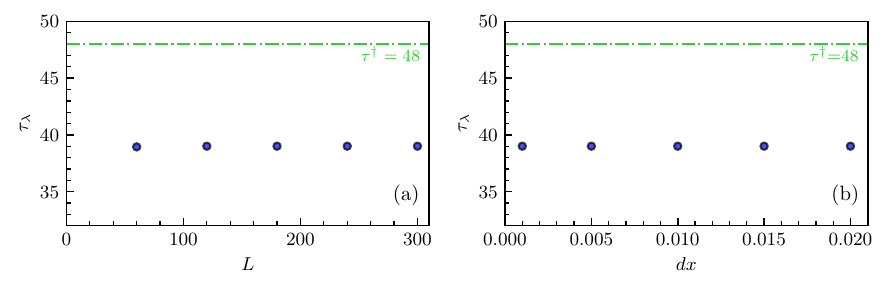}
    \caption{Numerically extracted rate of convergence for pushed fronts is not dependent on simulation parameters. We numerically extract $\tau^{\lambda}$ (blue circles) via a log-linear fit as described in Section IV. A. We compare $\tau^{\lambda}$ to the theoretically predicted $\tau^{\dagger}$ (dash-dotted, green) over varying domain sizes (a) and spatial resolutions (b). Simulation parameters across both panels are $t=500$, $dt=0.025$, $\delta = 10$. In panel (a), $dz=0.001$ while $L \in \{60,120,180,240,300\}$. In panel (b), $L=60$ while $dz \in \{0.001, 0.005, 0.01, 0.015, 0.02\}$. These results validate the robustness of the GBC method.}
    \label{fig:Ec_convergence}
\end{figure}

\begin{figure}
    \centering
    \includegraphics[width=\linewidth]{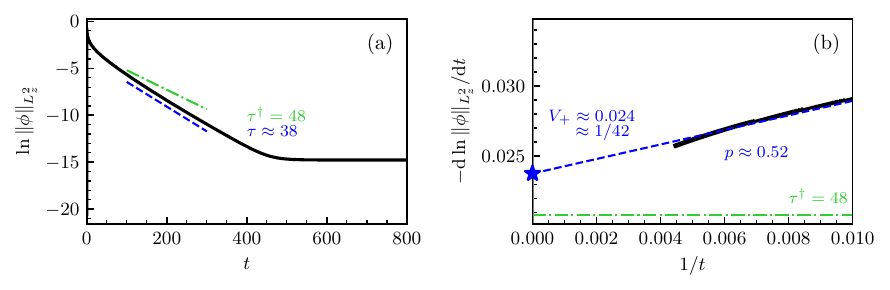}
    \caption{We use the $L^2$ norm of $\phi$ as the metric to extract the pushed front rate of convergence $\tau$ in two ways. Panel (a) depicts a log-linear fit used to extract numerically the time constant $\tau \approx 38$ (dashed, blue), compared against the theoretically predicted $\tau^\dagger = 48$ (dash-dotted, green). Panel (b) depicts the measured local rate against $1/t$, from which we extract the intercept (blue star) as $1/\tau = V_+ \approx 1/42$ and the slope $p \approx 0.52$.
    Simulation parameters are $t=1000$, $dt=0.025$, $L=60$, $dx=0.001$, $\delta = 10$.}
    \label{fig:V+_fit}
\end{figure}

With a naïve measurement of the exponential convergence, as depicted in \figsubref{fig:pushed_const}{e}, we find that $\tau=40$, which is close to the expected $48$. Moreover, as shown in Fig.~\ref{fig:Ec_convergence}, this measurement is robust to changes in the domain size and resolution, suggesting that discretization is not the cause of this discrepancy. Instead, we find an algebraic correction. This is done by relaxing the assumption of a single normal mode for $\eta(z,t)$ and writing $\eta(z,t) = e^{-v^\dagger z/2}\phi(z,t).$ Then $\phi(z,t)$ obeys
\begin{equation}
    \phi_t = \phi_{zz} - V(z)\phi.
\end{equation}
Far out along the front's edge, where $V(z)\to V_+$, this can be solved exactly to give
\begin{equation}
    \phi(z,t) = e^{-V_+t}\frac{1}{2\pi}\int_\mathbb{R} dk\, \tilde{\phi}_0(k)e^{ikz-k^2t},
\end{equation}
where $\tilde{\phi}_0(k)$ is the Fourier transform of the initial condition $\phi(z,0)$. If the integral is well-behaved (e.g. $\phi_0(z)=\delta(z)$), then we expect that, pointwise, 
\begin{equation}
    \phi(z,t)\propto t^{-1/2}e^{-V_+t}.
\end{equation}
The leading algebraic factor is the same as that of the heat kernel. For the $L^2$ norm, by Plancherel's theorem, we can work in Fourier space:
\begin{equation}
    \|\phi(\cdot,t)\|_{L^2_z}
    =
    \frac{1}{\sqrt{2\pi}}
    \|\widetilde{\phi}(\cdot,t)\|_{L^2_k}
    =
    e^{-V_+t}
    \left[
        \frac{1}{2\pi}
        \int_{\mathbb{R}}
        \left|
            \widetilde{\phi}_0(k)e^{-k^2t}
        \right|^2
        \,dk
    \right]^{1/2}.
\end{equation}
For $\phi_0(z)=\delta(z)$ (or, more generally, using Laplace's method),
we obtain
\begin{equation}
    \|\phi(\cdot,t)\|_{L^2_z}
    \propto t^{-1/4}e^{-V_+t}.
\end{equation}

For measuring a function like $t^{-p}e^{-V_+t}$, we can plot the derivative of the logarithm to obtain $-p/t - V_+$. Fig.~\ref{fig:V+_fit} indicates this algebraic correction roughly holds, and we numerically find $V_+\approx 1/42$ and $p\approx 0.52$. It is not clear why we find this value of $p$. Our initial data is nonlocal (relative to $u^{\dagger}$), but it does not decay with the power law that would give us this value. It is also not clear how the boundary forcing complicates the picture. Finally, we would like to connect this argument to the convergence of the weighted energy functional \eqref{eq:energy}. We leave these items to future study.

\section{Pushed--Pulled Transition Time}\label{app:pushed_pulled_transition_time}

We compute the theoretical pushed--pulled transition time $t_{\textrm{pp}}$ \eqref{eq:t_pp} for the quadratic-cubic nonautonomous F-KPP equation \eqref{eq:fisher-23} with the algebraic growth rate $a(t)=(\epsilon t)^{\alpha}$ with $\epsilon,\,\alpha > 0$ and a constant diffusion coefficient $d(t)=1$. We can write down the natural asymptotic velocity \eqref{eq:natural_asymptotic_velocity}, relevant for pulled fronts, and pushed asymptotic velocity \eqref{eq:pushed-speed}, as 
\begin{align}
v^{**}(t) &= (\alpha+2)\sqrt{\frac{{(\epsilon t)}^{\alpha}}{\alpha+1}}\\
    v^{\dagger}(t) &= 3\sqrt{\frac{(\epsilon t)^{\alpha} + 1}{2}} - \sqrt{\frac{2}{(\epsilon t)^{\alpha} + 1}}.
\end{align}
Integrating, we find the natural asymptotic position
\begin{equation}
    x^{**}(t) = \int_0^t v^{**}(s)\,ds  = 2\sqrt{\frac{\epsilon^\alpha}{\alpha+1}t^{\alpha+2}}
\end{equation}
and the pushed asymptotic position
\begin{equation}\label{eq:x_dagger}
    x^{\dagger}(t) = \int_0^t v^\dagger(s)\,ds = \frac{t}{\sqrt{2}}\left[\,_2F_1\left(-\frac{1}{2},\frac{1}{\alpha},1+\frac{1}{\alpha},-(\epsilon t)^{\alpha}\right) - 2\sqrt{\frac{(\epsilon t)^{\alpha}+1}{2}}\,_2F_1\left(1,\frac{1}{2}+\frac{1}{\alpha},1+\frac{1}{\alpha},-(\epsilon t)^{\alpha}\right) \right],
\end{equation}
where $_2F_1(a,b,c;z)$ is the hypergeometric function. To find $t_{\mathrm{pp}}$ in general, one then solves $x^{**}(t) - x^\dagger(t)=0$ numerically. However, when $\alpha = 1$, this expression simplifies to
\begin{equation}\label{eq:asymptotic_position_difference}
    x^{**}(t) - x^{\dagger}(t) = 2\sqrt{\frac{\epsilon}{2}t^3} - \sqrt{2}\left[\frac{1}{{\epsilon}}\left(1+2\sqrt{\frac{1}{1+\epsilon t}}\right)\left(-1+\sqrt{1+\epsilon t}\right) + t\left(-2\sqrt{\frac{1}{1+\epsilon t}}+\sqrt{1+\epsilon t}\right)\right].
\end{equation}
This expression vanishes when $t=1/\epsilon$, indicating that $t_{\mathrm{pp}}=1/\epsilon$, the same result as obtained from the time-frozen transition time $t^*$, i.e. $a(t_{\mathrm{pp}})=1$. Figures~\ref{fig:transition} and \ref{fig:transition_timing} compare these results with simulations.

\end{document}